\documentclass[reqno,a4paper,11pt]{article}
\pdfoutput=1
\usepackage{xcolor}

\usepackage{graphicx}
\usepackage[textwidth = 430 pt, textheight = 630 pt]{geometry}

\definecolor{MyDarkBlue}{rgb}{0.15,0.25,0.45}
\usepackage{epsfig,rotating}
\usepackage{amsfonts}
\usepackage{mathrsfs}
\usepackage{bbm}
\usepackage[normalem]{ulem}
\usepackage{booktabs}
\usepackage{enumerate}

\usepackage{booktabs}

\usepackage{latexsym}
\usepackage{amsthm}
\usepackage[all,knot]{xy}
\xyoption{arc}

\usepackage[utf8x]{inputenc}

\usepackage{hyperref}
\hypersetup{
hypertexnames=false,
colorlinks=true,
citecolor=MyDarkBlue,
linkcolor=MyDarkBlue,
urlcolor=MyDarkBlue,
pdfauthor={Meer Ashwinkumar, Lennart Schmidt, Meng-Chwan Tan},
pdftitle={Matrix Quantization of Classical Nambu Brackets and Super p-Branes},
pdfsubject={hep-th math-ph},
breaklinks=true
}
\usepackage{cite}

\usepackage{tikz}
\usetikzlibrary{matrix,cd,arrows}
\usepackage{mathtools}
\usepackage{tensor}
\usepackage[aligntableaux=center]{ytableau}
\usepackage[all,knot]{xy}
\xyoption{arc}

\let\fn\footnote
\renewcommand{\footnote}[1]{\linespread{1.1}\fn{#1}\linespread{1.29}}

\makeatletter\renewcommand{\section}{\@startsection
{section}{1}{\z@}{-3.5ex plus -1ex minus
    -.2ex}{2.3ex plus .2ex}{\bf }}
\makeatletter\renewcommand{\subsection}{\@startsection{subsection}{2}{\z@}{-3.25ex
plus -1ex minus
   -.2ex}{1.5ex plus .2ex}{\bf }}
\makeatletter\renewcommand{\subsubsection}{\@startsection{subsubsection}{3}{\z@}{-3.25ex
plus -1ex minus -.2ex}{1.5ex plus .2ex}{\it }}
\renewcommand{\thesection}{\arabic{section}}
\renewcommand{\thesubsection}{\arabic{section}.\arabic{subsection}}
\renewcommand{\@seccntformat}[1]{\@nameuse{the#1}.~~}
\renewcommand{\theequation}{\thesection.\arabic{equation}}
\providecommand*{\xhookrightfill@}{%
  \arrowfill@{\lhook\joinrel\relbar}\relbar\rightarrow
}
\providecommand*{\xhookrightarrow}[2][]{%
  \ext@arrow 0395\xhookrightfill@{#1}{#2}%
}
\makeatletter \@addtoreset{equation}{section}
\def\Ddots{\mathinner{\mkern1mu\raise\p@
\vbox{\kern7\p@\hbox{.}}\mkern2mu
\raise4\p@\hbox{.}\mkern2mu\raise7\p@\hbox{.}\mkern1mu}}
\usepackage[toc,page]{appendix}

\renewcommand{\thethm}{\thesection.\arabic{thm}}

\renewcommand{\appendices}{
\section*{Appendix}\label{appendices}\setcounter{subsection}{0}
\addcontentsline{toc}{section}{Appendix}
\setcounter{equation}{0}
\makeatletter
\renewcommand{\theequation}{\Alph{subsection}.\arabic{equation}}
\renewcommand{\thesubsection}{\Alph{subsection}}
\renewcommand{\thethm}{\Alph{subsection}.\arabic{thm}}
\@addtoreset{equation}{subsection}
\@addtoreset{thm}{subsection}
\makeatother
}

\newcommand{\FC}{\mathbbm{C}}
\newcommand{\RZ}{\mathbbm{Z}}
\newcommand{\NN}{\mathbbm{N}}
\newcommand{\unit}{\mathbbm{1}}
\newcommand{\dd}{\mathrm{d}}
\newcommand{\CA}{\mathcal{A}}
\newcommand{\CL}{\mathcal{L}}
\newcommand{\CO}{\mathcal{O}}
\newcommand{\CN}{\mathcal{N}}
\newcommand{\del}{\partial}
\newcommand{\asl}{\mathfrak{sl}}
\newcommand{\agl}{\mathfrak{gl}}
\newcommand{\frg}{\mathfrak{g}}
\newcommand{\quartic}[1]{\tensor[_4]{#1}{}}  
\newcommand{\tntic}[1]{\tensor[_{2n}]{#1}{}}
\newcommand{\quagl}{\quartic{\agl}}
\newcommand{\quasl}{\quartic{\asl}}
\newcommand{\tnagl}{\tntic{\agl}}
\newcommand{\tnasl}{\tntic{\asl}}
\newcommand{\Mod}[1]{\ (\mathrm{mod}\ #1)}
\newcommand{\eand}{{\qquad\mbox{and}\qquad}}
\newcommand{\acton}{\vartriangleright}

\begin{document}
\begin{titlepage}
\begin{flushright}
\phantom{NUS--21--02}
\end{flushright}
\vskip2.0cm
\begin{center}
{\LARGE \bf Matrix Quantization of Classical Nambu Brackets and Super $p$-Branes}
\vskip1.5cm
{\Large Meer Ashwinkumar$^{a}$, Lennart Schmidt$^b$, Meng{-}Chwan Tan$^b$}
\setcounter{footnote}{0}
\renewcommand{\thefootnote}{\arabic{thefootnote}}
\vskip1cm
{\em ${}^a$ Kavli Institute for the Physics and Mathematics of the Universe (WPI)\\
The University of Tokyo Institutes for Advanced Study\\
The University of Tokyo\\
Kashiwa, Chiba 277{-}8583, Japan}\\[0.5cm]
{\em ${}^b$ Department of Physics\\
National University of Singapore\\
2 Science Drive 3, Singapore 117551}\\[0.5cm]
{Email: {\ttfamily meer.ashwinkumar@ipmu.jp~,~phylen@nus.edu.sg~,~mctan@nus.edu.sg}}
\end{center}
\vskip1.0cm
\begin{center}
{\bf Abstract}
\end{center}
\begin{quote}
We present an explicit matrix algebra quantization of the algebra of volume-preserving diffeomorphisms of the $n$-torus. That is, we approximate the corresponding classical Nambu brackets using $\asl(N^{\lceil\tfrac{n}{2}\rceil},\FC)$-matrices equipped with the finite bracket given by the completely anti-symmetrized matrix product, such that the classical brackets are retrieved in the $N\rightarrow \infty$ limit. We then apply this approximation to the super $4$-brane in $9$ dimensions and give a regularized action in analogy with the matrix quantization of the supermembrane. This action exhibits a reduced gauge symmetry that we discuss from the viewpoint of $L_\infty$-algebras in a slight generalization to the construction of Lie $2$-algebras from Bagger--Lambert $3$-algebras.
\end{quote}
\end{titlepage}

\tableofcontents
\section{Introduction}
The dynamics of the classical membrane moving in $d$ dimensions are governed by the action~\cite{Dirac:1962sg,Collins:1976sg}
\begin{equation}
S = -\int\!\dd^3 x\, \sqrt{-g(X)}~,
\end{equation}
where $g(X)$ is the pullback of a spacetime metric to the worldvolume of the membrane. This is the direct analogue of the well-known Nambu--Goto action of the classical string. Similarly to the Green--Schwarz formulation~\cite{Green:1984fd} of the superstring, there exist supersymmetric extensions for the membrane action, which require an additional Wess--Zumino--Witten-term to ensure the presence of an appropriate $\kappa$-symmetry and, thus, the right number of fermionic degrees of freedom. This $\kappa$-symmetry is dependent on certain Fierz identities which are only satisfied in $d=4,5,7$ and $11$ dimensions~\cite{Achucarro:1987sg}.

In $11$ dimensions, this results in the supermembrane action\cite{Bergshoeff:1987cm}
\begin{equation}\label{eq:supermembrane_action}
\begin{aligned}
S_{\mathrm{GS}}&= - \int\!\dd^3 x\, \bigg(\sqrt{-g(X,\theta)} + \tfrac{i}{2}\epsilon^{ijk} \bar\theta \Gamma_{\mu\nu} \partial_i \theta\Big( \left(\partial_j X^\mu - i\bar\theta\Gamma^\mu \partial_j \theta\right)\partial_k X^\nu \\ 
&\phantom{{}={} - \int\!\dd^3\sigma \bigg(}- \tfrac13 \bar\theta \Gamma^\mu \partial_j \theta \bar\theta \Gamma^\nu \partial_k \theta\Big)\bigg)~.
\end{aligned}
\end{equation}
Unlike the superstring, this action does not allow for a consistent first quantization. Instead, de Wit et al.~\cite{deWit:1988ig} proposed a finite-$N$ regularization, in which the above action in the light-cone gauge is regularized by an appropriate $\asl(N,\FC)$-matrix gauge theory. This was based on approximating the algebra of area-preserving diffeomorphisms which correspond to the residual symmetries of the supermembrane after imposing the light-cone gauge. Prior work showed that such an approximation in terms of $\asl(N,\FC)$-matrices indeed exists for spherical~\cite{Hoppe:Diss} and toroidal~\cite{Fairlie:1988qd,Hoppe:1988qt,de1990area} membranes which was later extended to arbitrary topologies~\cite{Bordemann:1993zv}. The resulting $\asl(N,\FC)$-matrix gauge theory is the same as the one-dimensional reduction of the 10d super-Yang--Mills theory, which lies at the heart of the connection between matrix models and M-theory~\cite{Banks:1996vh} and has been used to study the supermembrane from a multitude of perspectives, such as investigating its groundstate, spectrum, vertex operators and even scattering amplitudes~\cite{deWit:1989sg,Yi:9704098,Froehlich:1998sg,Sethi:1997pa,Porrati:9708119,Halpern:9712133,Hoppe:2000tj,Dasgupta:0003280,Plefka:1997hm,Nicolai:1998ic,Dasgupta:2002iy}.

Interestingly, the action~\eqref{eq:supermembrane_action} is only a specific case of a more general type of actions that describe the dynamics of higher dimensional $p\,$-branes. Supersymmetric extensions of these are possible in a variety of low dimensions --- see~\cite{Achucarro:1987sg,Fiorenza:2013nha} for a complete list known as the ``old brane scan". 
Furthermore, the ``new brane scan" allows for additional tensor multiplet fields enlarging the class of examples to the full brane-bouquet of M- and string theory~\cite{Fiorenza:2013nha,Huerta:2017utu,Aganagic:1997zk}.

The immediate question this paper contends with is whether or not these higher-dimensional super $p$-branes can be quantized by a matrix theory in analogy to the supermem\-brane. Such a quantization would require a matrix regularization of the higher classical Nambu brackets
\begin{equation}
\left\{-,\dots,-\right\}: \CA^{\otimes p} \longrightarrow \CA~,~~~\left\{f_1,\dots, f_p\right\} = \epsilon^{r_1 \dots r_p} \partial_{r_1}f_1 \dots \partial_{r_p}f_p~,
\end{equation}
of volume-preserving diffeomorphisms (where $\mathcal{A}$ is a vector space of functions), just as the Poisson bracket of area-preserving diffeomorphisms was approximated by $\asl(N,\FC)$-matrices in the supermembrane case. 

Since their introduction in 1973~\cite{Nambu:1973qe}, Nambu brackets have been widely studied and have long been thought to be of relevance to the dynamics of extended objects~\cite{Takhtajan:1993vr,Hoppe:1996xp,Minic:1999js,Kerner:2000mr,Awata:1999dz,Minic:2002pd,Curtright:0212267,Ho:2016hob}. Furthermore, they also appear more generally in other contexts such as integrability~\cite{Bayen:1975sg,Chatterjee:1996sg,Gonera:2001sg} and fluid dynamics~\cite{Nevir:1993sg,Blender:2015sg}. 

One particular area of focus has been the question of quantization. Already in his original work, Nambu pointed out the difficulties in quantizing his proposed mechanics and there have subsequently been numerous approaches based on relaxing different constraints. The insistence on the identity
\begin{equation}
\left\{\left\{f_1,\dots,f_p\right\},g_1,\dots,g_{p-1}\right\} = \sum\limits_{i=1}^p \left\{f_1,\dots,\left\{f_i,g_1,\dots,g_{p-1}\right\},\dots,f_p\right\}~,
\end{equation}
widely known as the \textit{fundamental identity}, led to the proposal of various quantum Nambu brackets based on, e.g., partial tracing, non-associative products and cubic matrices~\cite{Hoppe:1996xp,Awata:1999dz,Kawamura:2002yz,Kawamura:2003sg,Ho:0701130,Bai:2014sg,Yoneya:2016wqw}. In contrast, Curtright and Zachos~\cite{Curtright:0212267} proposed to forego the fundamental identity in favour of the anti-symmetrized products in a matrix algebra $\frg$,
\begin{equation}
\left[-,\dots,-\right]:\frg^{\otimes p}\longrightarrow \frg~,~~~\left[A_1,\dots A_{p}\right] = \sum\limits_{\sigma\in S_p} \mathrm{sgn}(\sigma) A_{\sigma(1)}\dots A_{\sigma(p)}~,
\end{equation}
also sometimes referred to as \textit{Nambu--Heisenberg commutators}~\cite{Takhtajan:1993vr}\footnote{For $p=3$, this bracket had already been proposed in Nambu's original work~\cite{Nambu:1973qe}.}. Another approach is that of deformation quantization: In~\cite{Takhtajan:1993vr} a generalization of the Moyal bracket was proposed and in \cite{Curtright:2002sr,Gautheron:1996sg} similar brackets were introduced, all of which also fail to satisfy the fundamental identity. More recently, in~\cite{Mylonas:2012pg,Mylonas:2013jha} phase space deformation quantization was shown to lead to consistent quantizations of Nambu brackets, that do satisfy the fundamental identity, which was subsequently shown in ~\cite{Aschieri_2015} to imply deformation quantization of Nambu brackets in terms of tripoducts induced by non-associative star products. Further quantization prescriptions of Nambu-Poisson structures are discussed in~\cite{DeBellis:2010pf,DeBellis:2010sy} for different non-commutative geometries. Lastly, in~\cite{Dito:1996xr,Dito:1996hn} Zariski quantization was successfully used to find a quantum Nambu bracket that also does satisfy the fundamental identity but presents in a very different form from the more conventional quantization methods.

In this article we show that --- in the toroidal case --- utilizing the Nambu--Heisenberg commutators does allow for a matrix approximation of all higher Nambu brackets and, furthermore, we provide a regularized matrix theory with reduced gauge symmetry in the case of a super $(d=9, p=4)$-brane.  We treat this simple example because, as an element of the old brane scan, it does not carry any gauge or higher form fields and therefore is in close analogy with the supermembrane case.

This adds evidence to the proposal of~\cite{Curtright:0212267} that the Nambu--Heisenberg commutators should be the relevant quantum Nambu bracket. Additionally, such a quantization of Nambu brackets is thought to play a role in the understanding of M-theory as a matrix model. For example, one may be able to use it to better understand previously proposed 3-algebra models for M-theory~\cite{Sato:2014e,Sato:2009mf,Sato:2010ca,Sato:2011gi,Sato:2013mja,Sato:2013hh,Sato:2013hua} or it may play a role in better understanding the Basu--Harvey equations and their connection to membrane models~\cite{Basu:2004ed,Bagger:2012jb}. An interesting first step would be to extend this example to odd-dimensional super $p$-branes or even branes in the new brane scan, which, in principle, should follow along similar lines.

The paper is outlined as follows: Section~\ref{sec:supermembrane} summarizes relevant facts for the supermem\-brane. We refer to~\cite{deWit:1988ig} and useful reviews~\cite{Nicolai:1998ic,Dasgupta:2002iy} for more details.

In section~\ref{sec:nambu_brackets}, we provide a brief overview of classical Nambu brackets and their finite counterpart, the Nambu--Heisenberg commutators. The latter exhibit an even-odd dichotomy~\cite{Curtright:0212267}, which we discuss from the viewpoint of $L_\infty$-algebras. 

In section~\ref{sec:matrix_approximation}, we present the main result of the paper and derive an explicit approximation of the classical Nambu brackets both in the even and odd degree cases. Based on the previous approaches using cubic matrices, we use higher index objects to showcase the analogy to the case of Poisson brackets but ultimately show that the approximation can also be expressed using the language of ordinary matrices.

Finally, section~\ref{sec:four_brane_action} applies the previous result to write down a regularized action for a super $(d=9, p=4)$-brane that exhibits a reduced gauge symmetry. We also discuss the Lie-algebraic origin of this reduced symmetry inspired by the Lie-algebraic origin of Bagger--Lambert $3$-algebras.
 
The appendix provides explicit calculations and proofs for formulae and statements used throughout the main body of the paper.

\section*{Acknowledgments}
We would like to thank Grigorios Giotopoulos, Zoe Wyatt and Masahito Yamazaki for helpful comments and discussions. This work was supported by MOE Tier~2 grant R{-}144{-}000{-}396{-}112 [MA, LS, M-CT]; and World Premier International Research Center Initiative (WPI), MEXT, Japan and JSPS KAKENHI Grant-in-Aid for Scientific Research (No. 19H00689) [MA].

\section{Gauge-fixed super \texorpdfstring{$p$}{p}-brane actions and their matrix quantization}\label{sec:supermembrane}
We shall first review the relevant facts concerning the gauge-fixing and quantization of the Green--Schwarz supermembrane action, which underlies the BFSS matrix model. More details and further background material can be found in the reviews~\cite{Nicolai:1998ic,Dasgupta:2002iy}. Subsequently, we discuss how the gauge-fixing generalizes to higher dimensional super $p$-brane actions and pose the problem of their matrix quantization.

\subsubsection{The supermembrane}
Let us start with the supersymmetric Nambu--Goto membrane. The worldvolume coordinates will be denoted $\sigma^i=(\tau,\sigma^r)$ (where $i=0,1,2$ and $r=1,2$) and the spacetime coordinates are denoted $X^{\mu}$ (where $\mu=0,1,\dots,10$). The supermembrane worldvolume action is given by
\begin{equation}
\begin{aligned}
S_{\mathrm{GS}} &= - \int\!\dd^3 x\, \bigg(\sqrt{-g(X,\theta)} + \tfrac{i}{2}\epsilon^{ijk} \bar\theta \Gamma_{\mu\nu} \partial_i \theta\Big( \left(\partial_j X^\mu - i\bar\theta\Gamma^\mu \partial_j \theta\right)\partial_k X^\nu \\ 
&\phantom{{}={} - \int\!\dd^3\sigma \bigg(}- \tfrac13 \bar\theta \Gamma^\mu \partial_j \theta \bar\theta \Gamma^\nu \partial_k \theta\Big)\bigg)
\end{aligned}
\end{equation}
where $g(X,\theta)$ is the supersymmetric completion of the pullback of the spacetime metric, $G$, to the worldvolume, i.e.,
\begin{equation}\label{pbmet} g_{ij} = G_{\mu\nu}(\partial_i X^\mu -i \bar\theta\Gamma^\mu \partial_i\theta)(\partial_j X^\nu -i \bar\theta\Gamma^\nu\partial_j\theta),
\end{equation}
and the remaining term is the WZW-term required for $\kappa$-symmetry~\cite{Bergshoeff:1987cm}. In addition to the $\kappa$-symmetry, this action is also invariant under gauge transformations associated with reparametrizations of the worldvolume, which allows us to gauge-fix the action to a much simpler form. 

In light of this purpose, we shall define the lightcone coordinates $X^{\pm}=\frac{1}{\sqrt{2}}(X^0\pm X^1)$, so that $\mu=(+,-,a =2,\dots,10)$, and use the $\tau$-dependent reparametrization as well as the $\kappa$-symmetry to go to the lightcone gauge
\begin{equation}\label{gf}
\begin{aligned}
X^+ &=X^+_0+\tau~,\\
\Gamma^+\theta &=0~.
\end{aligned}
\end{equation}
As a result, the WZW-term immediately simplifies to
\begin{equation}
    \CL_{\mathrm{WZW}} = i\epsilon^{rs}\partial_r X^a \bar{\theta} \Gamma^-\Gamma_a \partial_s\theta~.
\end{equation}

Furthermore, the kinetic term can be simplified as follows. In the lightcone gauge the metric components take the form
\begin{subequations}
\begin{align}
\label{eq:g00}    g_{00} &= 2(\partial_0 X^- -i \bar{\theta} \Gamma^-\del_0\theta) + \partial_0\Vec{X}\cdot \del_0\Vec{X}~,\\
\label{eq:g0r}    g_{0r} &= \partial_r X^- -i \bar{\theta} \Gamma^-\del_r\theta + \partial_0 \Vec{X}\cdot \partial_r \Vec{X}~,\\
\label{eq:grs}   g_{rs} &= \partial_r \Vec{X} \cdot \partial_s \Vec{X}~,
\end{align}
\end{subequations}
where $\Vec{X}$ contains the components $(X^2,X^3,\dots X^{10})$. Elementary matrix manipulations yield the identity $\det g_{ij}=(g_{00}-g_{0r}g^{rs}g_{0s})\bar{g}$ which we can use --- together with most of the remaining reparametrization invariance to set $g_{0r}=0$ --- to obtain
\begin{equation}\label{eq:g_to_gbar}
g = g_{00}\bar{g}~,
\end{equation}
where $\bar{g}$ is the determinant of ${g}_{rs}$. Now, consider the momentum conjugate to $X^-$, i.e., 
\begin{equation}\label{eq:conj_mom_pplus}
    P^+ \equiv P^+_0 \sqrt{w(\sigma)} = \frac{\partial\CL}{\partial(\partial_0 X^-)} = \sqrt{-\frac{\bar{g}}{g_{00}}}~,
\end{equation} 
where we used the ansatz $P^+= P^+_0 \sqrt{w(\sigma)}$ since the conjugate momenta $P^+$ transforms as a density. Here, $\sqrt{w(\sigma)}$ can be seen as coming from a fiducial metric on a worldvolume-slice of the membrane with coordinates $\sigma^r$ and should not be confused with the metric $g_{ij}$ on the full worldvolume. 
Combining~\eqref{eq:g_to_gbar} and~\eqref{eq:conj_mom_pplus} enables us to rewrite the kinetic term as
\begin{equation}
 \CL_{\mathrm{kin}} = -\sqrt{-g(X,\theta)} =- \frac{\bar g}{P^+_0 \sqrt{w(\sigma)}}~.
\end{equation}

To further simplify, we split the kinetic term in half and rewrite in two different ways. On the one hand, we can introduce the Poisson bracket associated with the symplectic form induced by $w$, that is
\begin{equation}\label{nambu2}
    \left\{A,B\right\} = \frac{\epsilon^{rs}}{\sqrt{w(\sigma)}} \partial_r A\partial_s B~.
\end{equation}
As the determinant of a matrix $M$ with entries $m_{ij}$ is given by 
\begin{equation}\label{detid}
\det(M) = \frac{1}{n!} \varepsilon^{i_1\dots i_n} \varepsilon^{j_1\dots j_n} m_{i_1 j_1} \dots m_{i_n j_n}
\end{equation}
we can subsequently when considering~\eqref{eq:grs} write
\begin{equation}\label{expl}
    -\frac{\bar g}{P_0^+ \sqrt{w(\sigma)}} = - \frac{\sqrt{w(\sigma)}}{2P_0^+} \{X^a,X^b \}\{X_a,X_b \}~.
\end{equation}

On the other hand, using~\eqref{eq:g00} and~\eqref{eq:conj_mom_pplus} we can write
\begin{equation}
    -\frac{\bar g}{P_0^+ \sqrt{w(\sigma)}} = \sqrt{w(\sigma)}P^+_0 g_{00} = \sqrt{w(\sigma)}P^+_0 ( \partial_0\Vec{X}\cdot \del_0\Vec{X}-2i \bar{\theta} \Gamma^-\del_0\theta)~,
\end{equation}
where we dropped the total derivative term that will integrate out.

Altogether, we arrive at the gauge-fixed supermembrane action
\begin{equation}\label{eq:gfaction}
\begin{aligned}
S_{\mathrm{lc}} = \int d\tau \int d^2 \sigma \sqrt{w(\sigma)}\bigg(&\frac{P_0^+}{2}\del_{0}X^a\del_{0}X_a-iP_0^+\bar{\theta}\Gamma^-\del_{0}\theta- \frac{1}{4P_0^+} \{X^a,X^b \}\{X_a,X_b \} \\& +i\bar{\theta} \Gamma^-\Gamma_a \{X^a,\theta\}\bigg)~,
\end{aligned}
\end{equation}
which takes the form of a one-dimensional supersymmetric Yang--Mills theory with gauge field fixed to zero, if one regards the integration measure $\int d^2 \sigma \sqrt {w(\sigma)}$ as a trace.

The group of gauge transformations in this case can be deduced by noting that~\eqref{nambu2} generates field transformations 
\begin{equation}\label{eq:apd}
\delta X=\{\xi, X\}~,
\end{equation}
where $\xi(\sigma)$ is a function on the coordinates $\sigma^r$ that serves as a gauge parameter. These transformations correspond to area-preserving diffeomorphisms of the membrane.

To see this, note that a diffeomorphism $\sigma^r \rightarrow f^r(\sigma)= \sigma^r + v^r(\sigma)$ is area-preserving when $\partial_{r}v^r=0$. Using Hodge theory, the general solution of the latter equation can be decomposed into co-exact and harmonic vector fields. The harmonic forms on the torus are constant and, thus, have a vanishing bracket so that we can ignore them in the following.

The co-exact components can be written as
\begin{equation}
v^r(\sigma)=\frac{\epsilon^{rs}}{\sqrt{w(\sigma)}}\del_s \xi (\sigma),
\end{equation} 
which precisely give rise to the transformations~\eqref{eq:apd}. They represent the remaining invariance of the gauge-fixed action~\eqref{eq:gfaction} and act as the global gauge invariance.

We can promote to local gauge invariance by introducing an auxiliary field, $A$, that transforms as a gauge field, i.e., 
\begin{equation}
    \delta A = \del_{0} \xi +\{\xi, A \}~,
\end{equation}
and defining the covariant derivative $\mathrm{D} X= \del_{0} X -\{A,X\}$, via which we arrive at the action
\begin{equation}\label{eq:gauged_lc_action}
\begin{aligned}
S^{\mathrm{g}}_{\mathrm{lc}} = \int d\tau \int d^2 \sigma \sqrt{w(\sigma)}\bigg(&\frac{P_0^+}{2}\mathrm{D}X^a\mathrm{D}X_a-iP_0^+\bar{\theta}\Gamma^-\mathrm{D}\theta- \frac{1}{4P_0^+} \{X^a,X^b \}\{X_a,X_b \} \\& +i\bar{\theta} \Gamma^-\Gamma_a \{X^a,\theta\}\bigg)~.
\end{aligned}
\end{equation}
This is a one-dimensional supersymmetric Yang--Mills theory invariant under local gauge transformations
\begin{equation}
\delta X = \{\xi,X\} \eand \delta\theta = \{\xi,\theta\}~,
\end{equation}
where $\xi(\tau,\sigma)$ is now a $\tau$-dependent gauge parameter.

The regularization of this action that first appeared in~\cite{deWit:1988ig} is based on approximating the Lie algebra of these area-preserving diffeomorphisms. Suitable finite approximations have been found for any topology of the membrane~\cite{Bordemann:1993zv}. Here, we consider the simple case of $T^2$ for which we have the convenient basis of functions given by the eigenfunctions of the Laplacian, i.e.,
\begin{equation}
    Y_{(m_1,m_2)} = e^{2\pi i(m_1\sigma_1+m_2\sigma_2)}~,
\end{equation} 
where $m_1,m_2\in\RZ~$. Expanding the fields $\Vec{X}$ in terms of this basis as
\begin{equation}
\Vec{X}(\sigma)=\Vec{X}^{\Vec{m}}Y_{\Vec{m}}(\sigma).
\end{equation}
and plugging this into the Poisson bracket leads to 
\begin{equation}\label{eq:infinitealg}
    \left\{Y_{(m_1,m_2)},Y_{(n_1,n_2)}\right\} = (2\pi i)^2(m_1n_2-m_2n_1) Y_{(m_1+n_1,m_2+n_2)}~.
\end{equation}
Note that the Fourier mode $Y_{0,0}$ does not appear in the gauge-fixed membrane action, and we therefore exclude it in the following.

In order to regularize the action one introduces a cut-off $N$ and replaces~\eqref{eq:infinitealg} with the finite-dimensional Lie algebra $\mathfrak{sl}(N,\mathbb{C})$, where the Poisson bracket is replaced by the ordinary commutator of finite-dimensional matrices. For the torus, the appropriate replacement uses a specific basis of $\mathfrak{sl}(N,\mathbb{C})$ generated by the clock matrix $U$ and shift matrix $V$ given by
\begin{equation}\label{eq:clock_and_shift}
    U = \begin{pmatrix}  1 &&&& 0 \\ & \omega &&& \\ && \ddots && \\ &&& \ddots & \\ 0 &&&& \omega^{N-1} \end{pmatrix} \eand V = \begin{pmatrix} 0 & 1 &&& 0 \\ & 0 & 1 &&& \\ &&\ddots&\ddots & \\ &&&\ddots &1\\ 1 &&&& 0 \end{pmatrix}~,
\end{equation}
where $\omega = \exp{\tfrac{2\pi i}{N}}$ is a primitive $N^{\text{th}}$ root of unity. These matrices satisfy the important properties
\begin{equation}\label{eq:clock_and_shift_algebra}
    U^N = V^N = 1 \eand VU = \omega UV~,
\end{equation}
and, moreover, the products $U^{m_1}V^{m_2}$ are traceless for $(m_1,m_2)\in \RZ_N^2\setminus \{(0,0)\}$ yielding $N^2{-}1$ linearly independent matrices, whose linear combinations do indeed generate $\mathfrak{sl}(N,\mathbb{C})$.
Using this, we can approximate the eigenfunctions above as 
\begin{equation}\label{eq:membrane_matrix_replacement}
    Y_{(m_1,m_2)} \overset{N\to\infty}{\longleftarrow} T_{(m_1,m_2)} =-2\pi i N\omega^{\frac{m_1m_2}{2}} U^{m_1}V^{m_2}~,
\end{equation}
where $T^{\dagger}_{(m_1,m_2)}=-T_{(-m_1,-m_2)}$.
With the choice of prefactor given in~\eqref{eq:membrane_matrix_replacement} and with the use of the identities~\eqref{eq:clock_and_shift_algebra}, the commutator yields
\begin{equation}\label{eq:membrane_finite_bracket}
    \left[T_{(m_1,m_2)},T_{(n_1,n_2)} \right] = 2\pi i N(\omega^{\frac{m_1n_2-m_2n_1}{2}}-\omega^{-\frac{m_1n_2-m_2n_1}{2}}) T_{(m_1+n_1,m_2+n_2)}~.
\end{equation}
Expanding the root of unity then leads to
\begin{equation}
    \left[T_{(m_1,m_2)},T_{(n_1,n_2)} \right] = \left((2\pi i)^2 (m_1n_2-m_2n_1) +O\left(\tfrac{1}{N}\right)\right) T_{(m_1+n_1,m_2+n_2)}~,
\end{equation}
which in the large $N$ limit indeed approximates the structure constants~\eqref{eq:infinitealg} of the infinite-dimensional Poisson bracket. 

To quantize the supermembrane action we can now replace $\int\!\dd^2\sigma \sqrt{w(\sigma)}$ and $\{-,-\}$ in~\eqref{eq:gauged_lc_action} by the matrix trace and commutator, respectively, to arrive at the one-dimensional matrix gauge theory
\begin{equation}\label{eq:reg_membrane_action}
\begin{aligned}
S^{\mathrm{g}}_{\mathrm{reg.}} = \int\!\dd\tau\,\, \mathrm{Tr}\bigg(&\frac{P_0^+}{2}\mathrm{D}X^a\mathrm{D}X_a-iP_0^+\bar{\theta}\Gamma^-\mathrm{D}\theta- \frac{1}{4P_0^+} [X^a,X^b ][X_a,X_b]\\
& +i\bar{\theta} \Gamma^-\Gamma_a [X^a,\theta]\bigg)~.
\end{aligned}
\end{equation}
where, now, $\mathrm{D}X= \partial_0 X -\left[A,X\right]$ ($A$ being a connection on a principal $\mathsf{SL}(N,\mathbb{C})$-bundle over $\mathbb{R}$) and the local gauge transformations are given by
\begin{equation}
\delta X = \left[\xi,X\right] \eand \delta\theta = \left[\xi,\theta\right]~.
\end{equation}

\subsubsection{General super $p$-branes}
The above membrane quantization is what we want to emulate for higher super $p$-branes. To this end, note that  the gauge-fixing of the membrane action readily generalizes to higher dimensional super $p$-brane actions. Starting from the worldvolume action 
\begin{equation}
    S=-\int d^{p+1}x \bigg(\sqrt{-g(X,\theta)}+\CL_{\mathrm{WZW}}\bigg),
\end{equation}
and using the lightcone gauge~\eqref{gf}, one arrives at an action involving 
\textit{Nambu brackets} --- see section~\ref{sec:nambu_brackets} for more details --- given by
\begin{equation}\label{eq:pbrackets}
    \left\{X_1,\dots,X_p\right\} = \frac{\epsilon^{r_1\dots r_p}}{\sqrt{w(\sigma)}} \partial_{r_1}X_1\dots\partial_{r_p}X_p~,
\end{equation}
where $w(\sigma)$ is now the fiducial metric of a $p$-dimensional worldvolume slice, parametrized by  coordinates $\sigma^r$, where $r=1,\dots,p$. The simplifications due to the lightcone gauge follow in the same way as before with the only difference being the generalization of~\eqref{expl}, which for higher dimensional super $p$-branes reads as
\begin{equation}
    -\frac{\bar g}{P_0^+ \sqrt{w(\sigma)}} = -\frac{\sqrt{w(\sigma)}}{p! P_0^+} \{X^{a_1},X^{a_2},\dots,X^{a_p} \}\{X_{a_1},X_{a_2},\dots,X_{a_p} \}~,
\end{equation}
where $\bar{g}$ is now the determinant of $g_{rs}$ with $r,s=1,\dots,p$. The resulting gauge-fixed action  is
\begin{equation}\label{eq:lc_action_pbrane}
\begin{aligned}
S_{\mathrm{lc}} =\int d\tau \int d^2 \sigma \sqrt{w(\sigma)}\bigg(&\frac{P_0^+}{2}\del_{0}X^a\del_{0}X_a - i P_0^+ \bar\theta\Gamma^-\partial_0\theta + \CL_{\mathrm{WZW}}\\
&- \frac{1}{2}\frac{1}{p! P_0^+} \{X^{a_1},X^{a_2},\dots,X^{a_p} \}\{X_{a_1},X_{a_2},\dots,X_{a_p} \}\bigg).
\end{aligned}
\end{equation}
in complete analogy with~\eqref{eq:gfaction}. In order to achieve a quantized action analogous to~\eqref{eq:reg_membrane_action}, it remains to gauge the residual symmetries, find an appropriate finite regularization of the Nambu brackets and write down a corresponding matrix gauge theory. This is the core of the discussion in the following sections.

\section{Nambu brackets, Nambu--Heisenberg commutators and \texorpdfstring{$L_\infty$}{Linf}-algebras}\label{sec:nambu_brackets}
In~\cite{Nambu:1973qe}, Nambu suggested an elegant generalization of Hamiltonian mechanics to a 3-dimensional phase space using his classical 3-bracket
\begin{equation}
\left\{f,g,h\right\} = \epsilon^{\mu\nu\kappa} \partial_\mu f \partial_\nu g \partial_\kappa h~,
\end{equation}
in place of the ordinary Poisson bracket. Moreover, this procedure generalizes to phase spaces of arbitrary dimensions when using higher brackets of order $p$, i.e.
\begin{equation}\label{eq:classical_NB_form}
\left\{f_1,\dots, f_p\right\} = \epsilon^{r_1 \dots r_p} \partial_{r_1}f_1 \dots \partial_{r_p}f_p~.
\end{equation}
Since then, Nambu brackets have been widely studied, see e.g.~\cite{Ho:2016hob,Curtright:0212267} for brief reviews and more references, and are thought to be of particular relevance to the dynamics of extended objects~\cite{Takhtajan:1993vr,Hoppe:1996xp,Minic:1999js,Kerner:2000mr,Awata:1999dz,Minic:2002pd,Curtright:0212267,Ho:2016hob}.

First, note that these brackets~\eqref{eq:classical_NB_form} satisfy
\begin{enumerate}[(i)]
\item complete \emph{anti-symmetry}:
\begin{equation}
\left\{f_{\sigma(1)},\dots,f_{\sigma(p)}\right\} = \text{sgn}(\sigma) \left\{f_1,\dots,f_p\right\}~,
\end{equation}
\item the \emph{Leibniz rule}\label{renum:Leibniz}:
\begin{equation}
\left\{fg,f_2,\dots,f_p\right\} = f\left\{g,f_2,\dots,f_p\right\}+\left\{f,f_2,\dots,f_p\right\}g~,
\end{equation}
\item and the \emph{fundamental identity}\label{renum:FI}:
\begin{equation}
\left\{\left\{f_1,\dots,f_p\right\},g_1,\dots,g_{p-1}\right\} = \sum\limits_{i=1}^p \left\{f_1,\dots,\left\{f_i,g_1,\dots,g_{p-1}\right\},\dots,f_p\right\}~.
\end{equation}
\end{enumerate}

In light of this, Nambu brackets have subsequently been more abstractly defined~\cite{Takhtajan:1993vr} as multilinear maps on a ring $\CA$ of $C^\infty$-functions, that is
\begin{equation}
\left\{-,\dots,-\right\}:\CA^{\otimes p}\to\CA~,~~~(f_1,\dots,f_p) \mapsto \left\{f_1,\dots,f_p\right\}~,
\end{equation} 
which satisfy the above properties. 

For the purposes of this paper, however, we are only interested in the classical Nambu brackets~\eqref{eq:classical_NB_form}, as it is these that appear in the Lagrangians describing the various branes appearing in M- and string-theory.  It is an open question what potential finite replacement, or quantum Nambu bracket, should be used to play the same role as commutators do for the Poisson bracket. The most natural candidate is the completely anti-symmetrized product, also known as the \textit{Nambu--Heisenberg commutator}, that is
\begin{equation}\label{eq:anti_symm_prod_bracket}
\left[A_1,\dots A_{p}\right] = \sum\limits_{\sigma\in S_p} \mathrm{sgn}(\sigma) A_{\sigma(1)}\dots A_{\sigma(p)}~,
\end{equation}
which, however, fails to satisfy~\eqref{renum:Leibniz} and~\eqref{renum:FI}, except in the case of $p=2$, where the fundamental and Jacobi identities coincide. 

Consequently, the insistence on both the fundamental identity and Leibniz rule has led researchers to forego the Nambu--Heisenberg commutators and consider other possibilities in the search for an appropriate quantum Nambu bracket~\cite{Hoppe:1996xp,Awata:1999dz,Kawamura:2002yz,Kawamura:2003sg,Ho:0701130,Bai:2014sg,Yoneya:2016wqw}.

Here, we adopt a different viewpoint: While the fundamental identity is sometimes thought of as a generalization of the Jacobi identity, it is not unique in doing so and, more importantly, not the most natural. Similarly, while the Leibniz rule is required for consistency when viewing the bracket as providing the temporal derivative in multi-Hamiltonian dynamics, this is of no relevance for our purposes. Thus, we drop the insistence on these identities which allows us to work with the brackets~\eqref{eq:anti_symm_prod_bracket}, the consequences of which we discuss more closely in section~\ref{sec:four_brane_action}.

This approach follows the viewpoint of~\cite{Curtright:0212267,Curtright:0303088}, where it is shown that the brackets~\eqref{eq:anti_symm_prod_bracket} instead satisfy a different set of identities which can be thought of as encoding the associativity of the underlying operator product. Here, we add that these identities can be conveniently seen as particular cases of $L_\infty$-algebras which are well-known generalizations of Lie algebras and the Jacobi identity. As $L_\infty$-algebras enable the use of a standardized framework for gauge theories, see e.g.~\cite{Sati:2009sg,Jurco:2019bvp}, viewing Nambu--Heisenberg commutators in their context is also helpful in understanding the reduced gauge symmetry appearing in the quantized action for the super $(d=9,p=4)$-brane discussed in section~\ref{sec:four_brane_action}.

Let us quickly recall the relevant definitions: An \emph{$L_\infty$-algebra} $L$ consists of a $\RZ$-graded vector space $L = \bigoplus_{k\in \RZ} L_k$ together with a set of totally antisymmetric, multilinear maps or \emph{higher brackets} $\mu_i :  \wedge^i L \to L,$ $i\in \NN^+$, of degree~$2-i$, which satisfy the \emph{higher} or \emph{homotopy Jacobi identities}
\begin{equation}\label{eq:hom_rel}
 \sum_{i+j=m}\sum_{\sigma\in S_{i|j}}\chi(\sigma;a_1,\dots,a_{m})(-1)^{j}\mu_{j+1}(\mu_i(a_{\sigma(1)},\dots,a_{\sigma(i)}),a_{\sigma(i+1)},\dots,a_{\sigma(m)})=0
\end{equation}
where all $m\in\NN^+$ and $a_1,\dots,a_m\in L$ and where the second sum runs over all $(i,j)$-unshuffles $\sigma\in S_{i|j}$. An \emph{$n$-term $L_\infty$-algebra} is an $L_\infty$-algebra that is concentrated (i.e.~non-trivial only) in degrees $-n+1,\dots,0$. Here, an {\em unshuffle} $\sigma\in S_{i|j}$ is a permutation whose image consists of ordered tuples $\big(\sigma(1),\dots,\sigma(i)\big)$ and $\big(\sigma(i+1),\dots,\sigma(m)\big)$. Moreover, $\chi(\sigma; a_1,\dots,a_m)$ denotes the {\em graded antisymmetric Koszul sign} defined by the graded antisymmetrized products
\begin{equation}
a_1 \dots   a_m = \chi(\sigma;a_1,\dots,a_m)a_{\sigma(1)} \dots  a_{\sigma(m)}~,
\end{equation}
where any transposition involving an even element acquires a minus sign.

The simplest case of a $1$-term $L_\infty$-algebra is an ordinary Lie algebra and higher $n$-term $L_\infty$-algebras represent generalizations thereof. In particular, a $2$-term $L_\infty$-algebra is equivalent to a categorification of a Lie algebra to a Lie 2-algebra, where we relax the Jacobi identity to hold only up to a natural transformation~\cite{Baez:2003fs}. This is evident in the lowest few homotopy Jacobi relations, i.e.
\begin{equation}\label{eq:hJacobi_Lie_2}
\begin{aligned}
0&=\mu_1\left(\mu_1\left(a_1\right)\right)~,\\
0&= \mu_1\left(\mu_2\left(a_1,a_2\right)\right)-
\mu_2\left(\mu_1\left(a_1\right),a_2\right)+
(-1)^{\left| a_1\right|  \left| a_2\right| } \mu_2\left(\mu_1\left(a_2\right),a_1\right)~,\\
0 &=\mu_1\left(\mu_3\left(a_1,a_2,a_3\right)\right)-\mu_2\left(\mu_2\left(a_1,a_2\right),a_3\right)+(-1)^{|a_2| |a_3|} \mu_2(\mu_2(a_1,a_3),a_2)-\\
&~~~~-(-1)^{|a_1|(|a_2|+ |a_3|)} \mu_2(\mu_2(a_2,a_3),a_1)-
(-1)^{|a_1| |a_2|} \mu_3(\mu_1(a_2),a_1,a_3)+\\
&~~~~+
\mu_3(\mu_1(a_1),a_2,a_3)+
(-1)^{(|a_1|+|a_2|) |a_3|} \mu_3(\mu_1(a_3),a_1,a_2)~,
\end{aligned}
\end{equation}
where $a_i \in L$. These relations state that $\mu_1$ is a graded differential compatible with $\mu_2$, and $\mu_2$ is a generalization of a Lie bracket with the violation of the Jacobi identity controlled by $\mu_3$.

As mentioned in the introduction, the brackets~\eqref{eq:anti_symm_prod_bracket} exhibit an even-odd dichotomy: For $p=2n\in 2\NN^+$, they satisfy the identity~\cite{Curtright:0212267,Curtright:0303088}
\begin{equation}\label{eq:even_id}
\sum\limits_{\sigma \in S_{2p-1}} \text{sgn}(\sigma) \left[\left[A_{\sigma(1)},\dots,A_{\sigma(p)}\right],A_{\sigma(p+1)},\dots,A_{\sigma(2p-1)}\right] = 0~.
\end{equation}
Thus, letting $\frg$ denote the matrix algebra\footnote{In our case, $\frg$ will be given by $\asl(N^k,\FC)$ for some power $k$.} the $A_i$ are elements of, this forms a $(2p-3)$-term $L_\infty$-algebra of the form
\begin{equation}\label{eq:even_l_inf_form}
\begin{tikzcd}[outer sep=+2pt]
L = \frg[4{-}2p] \arrow[r,"\mu_1=0"] & \ast \dots \ast\arrow[r,"\mu_1=0"] & \frg[2-p]\arrow[r,"\mu_1=0"]  & \ast \dots\ast \arrow[r,"\mu_1=0"]& \frg
\end{tikzcd}
\end{equation}
together with the brackets
\begin{equation}
\begin{aligned}
\mu_p&: \frg \wedge \dots \wedge \frg \to \frg[2-p]~,~~~& \mu_p(A_1,\dots,A_p) &= \left[A_1,\dots,A_p\right]~,\\
\mu_p&: \frg[2-p]\wedge \frg \wedge \dots \wedge \frg \to \frg[4{-}2p]~,~~~& \mu_p(A_1,\dots,A_p) &= \left[A_1,\dots,A_p\right]~.\\
\end{aligned}
\end{equation}
Here, $\frg[k]$ denote the algebra $\frg$ with elements shifted in degree by $k$, which constitute the graded vector space underlying the $L_\infty$-algebra $L$, and $\ast$ denotes the trivial vector space. As there are only two non-trivial brackets most of the higher Jacobi relations~\eqref{eq:hJacobi_Lie_2} are satisfied trivially and we are left with the case where $i=j+1=p$. As all elements involved are of even degree, the Koszul sign $\chi(\sigma;a_1,\dots,a_n)$ simplifies to give the sign of the permutation $\text{sgn}(\sigma)$ and we are left with the identity~\eqref{eq:even_id}.

Similarly, for $p=2n-1\in 2\NN^+-1$ the brackets~\eqref{eq:anti_symm_prod_bracket} satisfy the slightly modified identity~\cite{Curtright:0212267,Curtright:0303088}
\begin{equation}\label{eq:odd_id}
\sum\limits_{\sigma \in S_{2p-1}} \text{sgn}(\sigma) \left[\left[A_{\sigma(1)},\dots,A_{\sigma(p)}\right],A_{\sigma(p+1)},\dots,A_{\sigma(2p-1)}\right] = p \left[A_1,\dots,A_{2p-1}\right]~.
\end{equation}
Again, this can be seen as forming a $(2p-2)$-term $L_\infty$-algebra of the form
\begin{equation}\label{eq:odd_linf}
\begin{tikzcd}[outer sep=+2pt]
L=\frg[3{-}2p]\arrow[r,"\mu_1=\mathrm{id}"] & \frg[4{-}2p]\arrow[r,"\mu_1=0"] & \ast \dots \ast\arrow[r,"\mu_1=0"] & \frg[2-p]\arrow[r,"\mu_1=0"] &\ast \dots\ast\arrow[r,"\mu_1=0"] & \frg
\end{tikzcd}
\end{equation}
together with the brackets
\begin{equation}
\begin{aligned}
&\mu_1: \frg[3{-}2p] \to \frg[4{-}2p]~,& \mu_1(A) &= A~,\\
&\mu_p: \frg \wedge \dots \wedge \frg \to \frg[2-p]~,& \mu_p(A_1,\dots,A_p) &= \left[A_1,\dots,A_p\right]~,\\
&\mu_p: \frg[2-p]\wedge \frg \wedge \dots \wedge \frg \to \frg[4{-}2p]~,& \mu_p(A_1,\dots,A_p) &= \left[A_1,\dots,A_p\right]~,\\
&\mu_{2p-1}: \frg \wedge \dots \wedge \frg \to \frg[3{-}2p]~,& \mu_{2p-1}(A_1,\dots,A_{2p-1}) &= \left[A_1,\dots,A_{2p-1}\right]~.
\end{aligned}
\end{equation}
As before, due to most brackets vanishing, the only remaining non-trivial higher Jacobi relation reduces to the relevant identity~\eqref{eq:odd_id}.

\section{Matrix quantization of classical Nambu brackets}\label{sec:matrix_approximation}
In this section, we present the main results of this paper and derive the matrix quantization of the classical Nambu brackets in the toroidal case. We start by briefly introducing the many-index objects we use and proceed to derive the matrix quantization for Nambu brackets of even degrees. For illustrative purposes, this is first done for the quartic case before moving on to the general case. The quantization of Nambu brackets of odd degrees is then constructed as a reduction of the even degree cases. Again for illustrative purposes, we start with the cubic case before moving on to the general case.

\subsection{Quartic and higher \texorpdfstring{$2n$}{2n}-tic matrices}
It is a reasonable possibility that many-index objects can be used to quantize Nambu brackets, just as matrices have been used to quantize the ordinary Poisson bracket. Indeed, as outlined above, there have been attempts to use three-index objects endowed with a triple product, that is, cubic matrices, in order to quantize the cubic Nambu bracket.

E.g., in~\cite{Awata:1999dz,Bai:2014sg} the use of cubic matrices allowed for various quantized Nambu brackets that satisfy the fundamental identity. However, these brackets rely on tracing over one of the cubic matrices making it difficult to envision an explicit realization of the cubic Nambu bracket relations in terms of cubic matrices analogous to the membrane quantization of~\cite{deWit:1988ig}. Furthermore, it is not clear which product should be used for cubic matrices and various ternary and binary relations have been proposed, some of which are non-associative. See~\cite{Ladra:2016sg} for a discussion of a large class of associative binary products for cubic matrices and~\cite{Kerner:0004031} for an overview of ternary relations in physics.

In this paper, we will make use of quartic, and subsequently higher $2n$-index, matrices instead. In contrast to cubic matrices, there exists a natural associative binary product for these that generalizes the ordinary product of matrices. Endowed with this product these higher matrices are a nice tool to draw analogies and make generalizations more apparent. Note, however, that this is just a bookkeeping device and can be fully realized using ordinary matrices of correspondingly larger size. Here, we list the definitions and basic properties that we will use.

A \textit{quartic matrix $A$ of size $N$} is a 4-index object $A_{a_1a_2a_3a_4}\!\in\!\FC$, $a_i\in\left\{1,\dots,N\right\}$, endowed with the binary product
\begin{equation}
(A \quartic{\star} B)_{a_1 a_2 a_3 a_4} = \sum\limits_{m_1, m_2} A_{a_1a_2m_1m_2}B_{m_2m_1a_3a_4}~.
\end{equation}
This forms a semigroup that we denote by $(\quartic{\mathsf{Mat}}(N,\FC)\equiv\quagl(N,\FC)\,,\,\quartic{\star})$. The above product is non-commutative and a quick check,
\begin{equation}
\begin{aligned}
((A\quartic{\star}B)\quartic{\star}C)_{a_1a_2a_3a_4}&= A_{a_1a_2m_3m_4}B_{m_4m_3m_1m_2}C_{m_2m_1a_3a_4}\\
&=A_{a_1a_2m_1m_2}B_{m_2m_1m_3m_4}C_{m_4m_3a_3a_4}\\
&=(A\quartic{\star}(B\quartic{\star}C))_{a_1a_2a_3a_4}~,
\end{aligned}
\end{equation}
where we implicitly sum over any repeated indices, shows that it is indeed associative. We have an identity element given by
\begin{equation}
(\quartic{\unit})_{a_1a_2a_3a_4} = \delta_{a_1a_4}\delta_{a_2a_3}
\end{equation}
and, in analogy, denote the group of invertible elements by $\left(\quartic{\mathsf{GL}}(N,\FC)\,,\,\quartic{\star}\right)$. Lastly, we also define a trace operation $\quartic{\mathrm{tr}}$ given by
\begin{equation}
\quartic{\mathrm{tr}}(A) = \sum\limits_{a_1, a_2} A_{a_1a_2a_2a_1}~,
\end{equation}
which, just like the ordinary trace, satisfies $\quartic{\mathrm{tr}}(A\quartic{\star}B)=\quartic{\mathrm{tr}}(B\quartic{\star}A)$. As noted above, this can also be realized by ordinary matrices of size $N^2$. More specifically, there is a semigroup isomorphism given by
\begin{equation}\label{eq:quartic_semigroup_iso}
\begin{aligned}
\Phi:&& \quartic{\mathsf{Mat}}(N,\FC) &\to \mathsf{Mat}(N^2,\FC)~,\\
&& \Big\{A_{a_1a_2a_3a_4}\Big\}_{a_i=0}^{N-1} &\mapsto \Big\{A_{\phi_0(i)\phi_1(i)\phi_1(j)\phi_0(j)}\Big\}_{i,j=0}^{N^2{-}1}~,
\end{aligned}
\end{equation}
where we use the shorthand
\begin{equation}
\phi_k(x) = \lfloor\tfrac{x}{N^k}\rfloor\!\! \Mod{N}~.
\end{equation}
It is an easy exercise to check that this indeed maps the above product, identity and trace to the ordinary matrix product, identity and trace.

More generally, we define \textit{$2n$-tic matrices of size $N$} as $2n$-index objects $A_{a_1\dots a_{2n}}\!\in\!\FC$, $a_i\in\left\{1,\dots,N\right\}$, endowed with the binary product
\begin{equation}
(A \tntic{\star} B)_{a_1\dots a_{2n}} = \sum\limits_{m_i} A_{a_1\dots a_n m_1\dots m_n} B_{m_n\dots m_1 a_{n+1}\dots a_{2n}}~.
\end{equation}
Again, this product is non-commutative and associative forming a semigroup denoted by $\left(\tntic{\mathsf{Mat}}(N,\FC)\equiv\tnagl(N,\FC)\,,\,\tntic{\star}\right)$. The identity element is now given by
\begin{equation}
(\tntic{\unit})_{a_1\dots a_{2n}} = \delta_{a_1 a_{2n}} \dots \delta_{a_n a_{n+1}}~,
\end{equation}
and the subgroup of invertible elements is denoted $\left(\tntic{\mathsf{GL}}(N,\FC)\,,\,\tntic{\star}\right)$. Finally, the trace is given by
\begin{equation}
\tntic{\mathrm{tr}}(A) = \sum\limits_{a_i} A_{a_1\dots a_n a_n \dots a_1}~,
\end{equation}
while the isomorphism to ordinary matrices is realized via
\begin{equation}\label{eq:gen_semigroup_iso}
\begin{aligned}
\Phi:&& \tntic{\mathsf{Mat}}(N,\FC) &\to \mathsf{Mat}(N^n,\FC)~,\\
&& \Big\{A_{a_1a_2a_3a_4}\Big\}_{a_i=0}^{N-1} &\mapsto \Big\{A_{\phi_0(i)\phi_1(i)\dots\phi_{n-1}(i)\phi_{n-1}(j)\dots\phi_1(j)\phi_0(j)}\Big\}_{i,j=0}^{N^n-1}~.
\end{aligned}
\end{equation}

\subsection{Matrix quantization of Nambu brackets for \texorpdfstring{$\mathbf{p\in 2\NN^+}$}{p in 2N}}
As discussed in section~\ref{sec:supermembrane}, there exists an explicit matrix quantization of the algebra of area-preserving diffeomorphisms for general topologies. For toroidal membranes, this explicit realization makes use of the clock and shift matrices~\eqref{eq:clock_and_shift} in order to approximate the Poisson bracket~\eqref{eq:infinitealg}.

Our aim is to generalize this to higher Nambu brackets: The cubic Nambu bracket, 
\begin{equation}
\left\{\varphi_1,\varphi_2,\varphi_3\right\} = \frac{\epsilon^{rst}}{\sqrt{\omega(\sigma)}} \partial_r \varphi_1 \partial_s \varphi_2 \partial_t \varphi_3~,
\end{equation}
being the next higher bracket, would be the most natural candidate and while, as discussed in section~\ref{sec:nambu_brackets}, it is an open question which finite bracket should be used to replace the Nambu bracket, the analogy with the Poisson bracket in the membrane case suggests to use the completely anti-symmetrized product
\begin{equation}\label{eq:cubic_antisymm_product_bracket}
\left[A,B,C\right] = ABC + BCA + CAB - ACB - CBA - BAC
\end{equation}
for our purposes.

Unfortunately, using this, it turns out to be infeasible to find an analogous, explicit matrix representation. Note that the anti-symmetrized product~\eqref{eq:cubic_antisymm_product_bracket} only satisfies the more complicated $L_\infty$-algebra relation~\eqref{eq:odd_id} that includes an ``inhomogeneity"~\cite{Curtright:0212267}. This suggests that it may be easier to focus on the Nambu--Heisenberg commutators that satisfy the simpler $L_\infty$-algebra relations~\eqref{eq:even_id} that do not include such a term. Thus, we will focus on Nambu brackets of order $p=2n$ corresponding to even-dimensional branes.

\subsubsection{The quartic case}
We start by considering the case $p=4$ $(n=2)$, that is, we consider the quartic Nambu bracket
\begin{equation}
\left\{\varphi_1,\varphi_2,\varphi_3,\varphi_4\right\} = \frac{\epsilon^{rstu}}{\sqrt{\omega(\sigma)}}\partial_r\varphi_1\partial_s\varphi_2\partial_t\varphi_3\partial_u\varphi_4~.
\end{equation}
For a toroidal 4-brane the eigenfunctions of the Laplace--Beltrami operator are given by 
\begin{equation}\label{ymforlater}
Y_{\vec{m}} = Y_{(m_1,m_2,m_3,m_4)} \coloneqq e^{2\pi i(m_1\sigma_1 + m_2\sigma_2 + m_3\sigma_3 + m_4\sigma_4)}~,
\end{equation}
where $\vec{m}\in\RZ^4$ and $\sigma_i \in \mathbbm{T}^4 = \left[0,1\right]^4$. For a flat torus we have $\sqrt{\omega(\sigma)}= 1$ leading to
\begin{equation}\label{eq:quartic_nambu_bracket}
\left\{Y_{\vec{m}},Y_{\vec{n}},Y_{\vec{k}},Y_{\vec{l}}\,\right\} =\left(2\pi\right)^4 \epsilon^{rstu} m_r n_s k_t l_u Y_{\vec{m}+\vec{n}+\vec{k}+\vec{l}}~,
\end{equation}
which should be approximated by an appropriate higher analogue of the shift and clock matrices together with the finite bracket
\begin{equation}\label{eq:finite_four_bracket}
\begin{aligned}
\left[A,B,C,D\right] &= ABCD - ABDC - ACBD + ACDB + ADBC - ADCB\\
&\phantom{{}={}} - BACD + BADC + BCAD - BCDA - BDAC + BDCA\\
&\phantom{{}={}} + CABD - CADB - CBAD + CBDA + CDAB - CDBA \\
&\phantom{{}={}}- DABC + DACB + DBAC - DBCA - DCAB + DCBA~.
\end{aligned}
\end{equation}

In the membrane case, forming the appropriate commutator identity~\eqref{eq:membrane_finite_bracket} crucially relies on the algebraic relations~\eqref{eq:clock_and_shift_algebra} satisfied by the clock and shift matrices. We, thus, propose an obvious generalization to this algebra by requiring the relations
\begin{equation}\label{eq:quartic_shift_and_clock_algebra}
\begin{gathered}
U_1^N=U_2^N=V_1^N=V_2^N = 1~,~~~V_1 U_1 = \omega U_1 V_1~,~~~ V_2 U_2 = \omega U_2 V_2~,\\
\left[U_1,U_2\right]=\left[U_1,V_2\right]=\left[U_2,V_1\right]=\left[V_1,V_2\right]=0~,
\end{gathered}
\end{equation}
where again $\omega = \exp \tfrac{2\pi i}{N}$ is a primitive $N^{\mathrm{th}}$ root of unity. Note that it is impossible to satisfy these relations using $(N\times N)$-matrices and, thus, we use quartic matrices $U_i, V_i \in \quagl(N,\FC)$ of size $N$ instead. As outlined above, this is equivalent to using $N^2\times N^2$-matrices but, as it keeps expressions simpler and the analogy more clear, we stick to quartic matrices for now and provide expressions for ordinary but larger matrices only at the end of this section. We also suppress the quartic and higher products $\quartic{\star}$ and $\tntic{\star}$ here and in the following as this can be inferred from context.

The reason underlying the fact that the clock and shift matrices $U$ and $V$ commute up to a factor of $\omega$ is that multiplication by $V$ from the left/right corresponds to shifting all columns right/all rows down, respectively. For the specific form of $U$, the difference is equivalent to multiplying by $\omega$.

Quartic matrices have two additional indices, so there are two more directions in which we can shift, allowing for two independent clock-and-shift pairs according to~\eqref{eq:quartic_shift_and_clock_algebra}. Explicitly, this straightforward generalization is given by quartic matrices $U_1,U_2,V_1,V_2 \in \quagl(N,\FC)$ with entries given by
\begin{equation}\label{eq:quartic_matrix_rep}
\begin{aligned}
(U_1)_{a_1a_2a_3a_4} &= \delta_{a_1a_4}\delta_{a_2a_3} \omega^{a_1{-}1}~,\\
(U_2)_{a_1a_2a_3a_4} &= \delta_{a_1a_4}\delta_{a_2a_3} \omega^{a_2{-}1}~,\\
(V_1)_{a_1a_2a_3a_4} &= \delta^N_{a_1+1,a_4}\delta_{a_2a_3}~,\\
(V_2)_{a_1a_2a_3a_4} &= \delta_{a_1a_4}\delta^N_{a_2+1,a_3}~,\\
\end{aligned}
\end{equation}
where $a_i \in \left\{0,\dots,N-1\right\}$ and we use the shorthand $\delta^N_{ij} \coloneqq \delta_{i \Mod{N}\,,\,j\Mod{N}}$. A straightforward calculation shows that these quartic matrices indeed satisfy the relations given in~\eqref{eq:quartic_shift_and_clock_algebra}, see appendix~\ref{app:explicit_calculations} for details.

In analogy with~\eqref{eq:membrane_matrix_replacement}, we can now assign quartic matrices to the basis functions according to
\begin{equation}\label{eq:quartic_matrix_basis}
Y_{(m_1,m_2,m_3,m_4)} \overset{N\to\infty}{\longleftarrow} T_{(m_1,m_2,m_3,m_4)} = \left(i\pi N \right)^{\tfrac{2}{3}} \omega^{\frac{m_1m_2+m_3m_4}{2}} U_1^{m_1}V_1^{m_2}U_2^{m_3}V_2^{m_4}~.
\end{equation}
Similarly to the clock and shift matrices, linear combinations of $U_1^{m_1}V_1^{m_2}U_2^{m_3}V_2^{m_4}$ generate all of $\quagl(N,\FC)$. This can be seen from the
general form
\begin{equation}\label{eq:explicit_product_quartic}
(U_1^{m_1}V_1^{m_2}U_2^{m_3}V_2^{m_4})_{a_1a_2a_3a_4} = \delta^N_{a_1+m_2,a_4}\delta^N_{a_2+m_4,a_3} \omega^{m_1(a_1{-}1)+m_3(a_2{-}1)}~.
\end{equation}

That is, we have $N^2$ quartic matrices of the form $U_1^{m_1}U_2^{m_3}$ that only have entries on the ``diagonal plane" defined by $\delta_{a_1a_4}\delta_{a_2a_3}$, are linearly independent and, therefore, span the set of such diagonal quartic matrices, just as powers of the clock matrix do for diagonal matrices. Furthermore, just as the shift matrix shifts to off-diagonals so do the quartic matrices $V_1$ and $V_2$ shift the diagonal plane and for each off-diagonal plane we get another set of $N^2$ linearly independent matrices. Thus, we have a complete basis for $\quagl(N,\FC)$.

As before, we exclude the identity $T_{(0,0,0,0)}=(i\pi N)^{2/3}\quartic{\unit}$ as it does not contribute to any action principles and are left with traceless quartic matrices which, in analogy with ordinary matrices, we denote by $\quasl(N,\FC) \coloneqq \left\{\, A \in \quagl(N,\FC)\mid \quartic{\mathrm{tr}}(A)=0\,\right\}$. With the phase factor in~\eqref{eq:quartic_matrix_basis} we have $\tensor*{T}{^\dagger_{\vec{m}}}=(-1)^{2/3}\tensor*{T}{_{-\vec{m}}}$ which also follows directly from the explicit form~\eqref{eq:explicit_product_quartic} and where transposing is understood to be given by reversing the order of the indices. 

Given this basis of $\quasl(N,\FC)$, we can simplify the terms appearing in the bracket~\eqref{eq:finite_four_bracket} by repeatedly using the basic relations~\eqref{eq:quartic_shift_and_clock_algebra} leading to
\begin{equation}
\begin{aligned}\label{eq:quartic_basis_product}
\tensor{T}{_{\vec{m}}}\tensor{T}{_{\vec{n}}}\tensor{T}{_{\vec{k}}}\tensor{T}{_{\vec{l}}} &= \left(i\pi N\right)^2 \omega^{\tfrac12\left(m_1(-n_2-k_2-l_2)+n_1(m_2-k_2-l_2)+k_1(m_2+n_2-l_2)+l_1(m_2+n_2+k_2)\right)}\\
&\phantom{{}={}} \times \omega^{\tfrac12\left(m_3(-n_4-k_4-l_4)+n_3(m_4-k_4-l_4)+k_3(m_4+n_4-l_4)+l_3(m_4+n_4+k_4)\right)}\\
&\phantom{{}={}} \times T_{\vec{m}+\vec{n}+\vec{k}+\vec{l}}\\
&\eqqcolon \left(i\pi N\right)^2 \omega^{e_4(\vec{m},\vec{n},\vec{k},\vec{l})}T_{\vec{m}+\vec{n}+\vec{k}+\vec{l}}~.
\end{aligned}
\end{equation}
Plugging this into the full bracket yields
\begin{equation}
\begin{aligned}
\left[T_{\vec{m}},T_{\vec{n}},T_{\vec{k}},T_{\vec{l}}\,\right] &= \left(i\pi N\right)^2 \left(\omega^{e_4(\vec{m},\vec{n},\vec{k},\vec{l})}-\omega^{e_4(\vec{m},\vec{n},\vec{l},\vec{k})}-\omega^{e_4(\vec{m},\vec{k},\vec{n},\vec{l})}+\omega^{e_4(\vec{m},\vec{k},\vec{l},\vec{n})}\right.\\
&+\omega^{e_4(\vec{m},\vec{l},\vec{n},\vec{k})}-\omega^{e_4(\vec{m},\vec{l},\vec{k},\vec{n})}-\omega^{e_4(\vec{n},\vec{m},\vec{k},\vec{l})}+\omega^{e_4(\vec{n},\vec{m},\vec{l},\vec{k})}\\
&+\omega^{e_4(\vec{n},\vec{k},\vec{m},\vec{l})}-\omega^{e_4(\vec{n},\vec{k},\vec{l},\vec{m})}-\omega^{e_4(\vec{n},\vec{l},\vec{m},\vec{k})}+\omega^{e_4(\vec{n},\vec{l},\vec{k},\vec{m})}\\
&+\omega^{e_4(\vec{k},\vec{m},\vec{n},\vec{l})}-\omega^{e_4(\vec{k},\vec{m},\vec{l},\vec{n})}-\omega^{e_4(\vec{k},\vec{n},\vec{m},\vec{l})}+\omega^{e_4(\vec{k},\vec{n},\vec{l},\vec{m})}\\
&+\omega^{e_4(\vec{k},\vec{l},\vec{m},\vec{n})}-\omega^{e_4(\vec{k},\vec{l},\vec{n},\vec{m})}-\omega^{e_4(\vec{l},\vec{m},\vec{n},\vec{k})}+\omega^{e_4(\vec{l},\vec{m},\vec{k},\vec{n})}\\
&\left.+\omega^{e_4(\vec{l},\vec{n},\vec{m},\vec{k})}-\omega^{e_4(\vec{l},\vec{n},\vec{k},\vec{m})}-\omega^{e_4(\vec{l},\vec{k},\vec{m},\vec{n})}+\omega^{e_4(\vec{l},\vec{k},\vec{n},\vec{m})}\right)T_{\vec{m}+\vec{n}+\vec{k}+\vec{l}}~.
\end{aligned}
\end{equation}
Now, expanding the root of unity leads to a cancellation of not only the constant but also the linear terms resulting in
\begin{equation}\label{eq:quartic_bracket_approx}
\begin{aligned}
\left[T_{\vec{m}},T_{\vec{n}},T_{\vec{k}},T_{\vec{l}}\right] &= \left(i\pi N\right)^2\left(4(2\pi i)^2\epsilon^{rstu} \frac{m_r n_s k_t l_u}{N^2} + \CO\left(\tfrac{1}{N^3}\right)\right)T_{\vec{m}+\vec{n}+\vec{k}+\vec{l}}\\
&= \Big((2\pi)^4\epsilon^{rstu} m_r n_s k_t l_u + \CO\left(\tfrac{1}{N}\right)\Big)T_{\vec{m}+\vec{n}+\vec{k}+\vec{l}}~,
\end{aligned}
\end{equation}
which for $N \longrightarrow \infty$ recovers the quartic Nambu bracket relations~\eqref{eq:quartic_nambu_bracket} as desired.

\subsubsection{The general case}
The above construction lends itself to a straightforward generalization to Nambu brackets of even order $p=2n$. That is, we have the bracket
\begin{equation}
\left\{\varphi_1,\dots,\varphi_{2n}\right\} = \frac{\epsilon^{r_1\dots r_{2n}}}{\sqrt{\omega(\sigma)}}\partial_{r_1}\varphi_1\dots\partial_{r_{2n}}\varphi_{2n}~,
\end{equation}
and for a toroidal brane in $p$ dimensions the eigenfunctions of the Laplace--Beltrami operator
\begin{equation}
Y_{\vec{m}_i} = Y_{(m_{i,1},\,\dots\,,m_{i,2n})} \coloneqq e^{2\pi i \sum\limits_{j=1}^{2n} m_{i,j} \sigma_j}~,
\end{equation}
where $\vec{m}_i\in\RZ^{2n}$ and $\sigma_j \in \mathbbm{T}^{2n} = \left[0,1\right]^{2n}$. Again, with $\sqrt{\omega(\sigma)}=1$, this leads to the bracket
\begin{equation}\label{eq:tntic_Nambu_bracket}
\left\{Y_{\vec{m}_1},\dots,Y_{\vec{m}_{2n}}\right\} = \left(2\pi i\right)^{2n}\epsilon^{r_1\dots r_{2n}} m_{1,r_1}\dots m_{2n,r_{2n}} Y_{\vec{m}_1+\,\dots\, + \vec{m}_{2n}}~,
\end{equation}
which we want to regularize using the finite Nambu--Heisenberg commutator
\begin{equation}\label{eq:tntic_finite_bracket}
\left[A_1,\dots A_{2n}\right] = \sum\limits_{\sigma\in S_{2n}} \mathrm{sgn}(\sigma) A_{\sigma(1)}\dots A_{\sigma(2n)}~.
\end{equation}
The algebra relations generalize to
\begin{equation}\label{eq:tntic_shift_and_clock_algebra}
\begin{gathered}
U_i^N=V_i^N= 1~,~~~V_i U_i = \omega U_i V_i~,\\
\left[U_i,U_j\right]=\left[U_i,V_j\right]=\left[V_i,V_j\right]=0~,
\end{gathered}
\end{equation}
where $i\neq j \in\left\{1,\dots,n\right\}$ and $\omega = \exp \tfrac{2\pi i}{N}$ is again a primitive $N^\mathrm{th}$ root of unity. These can now be satisfied by $2n$-tic matrices $U_i, V_i \in \tnagl(N,\FC)$ whose explicit expressions are given by
\begin{equation}\label{eq:gen_2n_matrices}
\begin{aligned}
(U_i)_{a_1\dots a_{2n}} &= \delta_{a_1a_{2n}}\dots\delta_{a_n a_{n+1}} \omega^{a_i-1}~,\\
(V_i)_{a_1\dots a_{2n}} &= \delta_{a_1a_{2n}}\dots\delta^N_{a_i+1,a_{2n-i+1}}\dots\delta_{a_n a_{n+1}}~,
\end{aligned}
\end{equation}
where, again, $a_i \in \{0,\dots,N-1\}$ and we use the shorthand $\delta^N_{ij} \coloneqq \delta_{i \Mod{N}\,,\,j\Mod{N}}$. As before, straightforward calculation show that these indeed satisfy~\eqref{eq:tntic_shift_and_clock_algebra}, see appendix~\ref{app:explicit_calculations} for details. 

The basis functions can be assigned to the $2n$-tic matrices
\begin{equation}
\begin{aligned}
Y_{(m_{i,1},\dots,m_{i,2n})} \overset{N\to\infty}{\longleftarrow} T_{(m_{i,1},\dots,m_{i,2n})} &= \left(\tfrac{(-2\pi iN)^n}{n!^2}\right)^{\tfrac{1}{2n-1}} \omega^{\frac{\sum\limits_{j=1}^{n}m_{i,2j-1}m_{i,2j}}{2}} \\
&\qquad \times U_1^{m_{i,1}}V_1^{m_{i,2}}\dots U_n^{m_{i,2n-1}}V_n^{m_{i,2n}}~.
\end{aligned}
\end{equation}
whose phase factor now leads to $\tensor{T}{^\dagger_{\vec{m}_i}} =(-1)^{n/(2n-1)} \tensor{T}{_{-\vec{m}_i}}$ and which, by the same argument as in the quartic case, generate all of $\tnagl(N,\FC)$. As before, they therefore form a basis for $\tnasl(N,\FC) \coloneqq \left\{\, A \in \tnagl(N,\FC) \mid \tntic{\mathrm{tr}}(A)=0 \,\right\}$ when excluding the identity $T_{\vec{0}} = (\tfrac{(-2\pi iN)^n}{n!^2})^{1/
(2n-1)} \tntic{\unit}$. 

A $2n$-fold product of this basis is of the form
\begin{equation}
T_{\vec{m}_1} \dots T_{\vec{m}_{2n}} = \tfrac{(-2\pi iN)^n}{n!^2} \omega^{e_{2n}(\vec{m}_1,\dots,\vec{m}_{2n})} T_{\vec{m}_1 + \dots + \vec{m}_{2n}}~,
\end{equation}
where repeated use of the algebra relations~\eqref{eq:tntic_shift_and_clock_algebra} can be used to show that the exponent is given by
\begin{equation}\label{eq:gen_exponent}
e_{2n}(\vec{m}_1,\dots,\vec{m}_{2n}) = \frac12\sum\limits_{j=1}^n \sum\limits_{i=1}^{2n}m_{i,2j-1}\left(\sum\limits_{k=1}^{i-1}m_{k,2j} - \sum\limits_{k=i+1}^{2n}m_{k,2j}\right)~.
\end{equation}

Plugging this into the Nambu--Heisenberg commutator~\eqref{eq:tntic_finite_bracket} subsequently yields
\begin{equation}\label{eq:general_formula_finite_bracket}
\left[T_{\vec{m}_1}, \dots, T_{\vec{m}_{2n}}\right] = \tfrac{(-2\pi iN)^n}{n!^2}\sum\limits_{\sigma\in S_{2n}} \mathrm{sgn}(\sigma) \omega^{e_{2n}(\vec{m}_{\sigma(1)},\dots,\vec{m}_{\sigma(2n)})} T_{\vec{m}_1 + \dots + \vec{m}_{2n}}~,
\end{equation}
and it remains to be shown that after expanding the root of unity this correctly recovers~\eqref{eq:tntic_Nambu_bracket}. Indeed, it follows from the symmetry properties of~\eqref{eq:gen_exponent} that we have
\begin{equation}\label{eq:gen_expanded_bracket}
\begin{aligned}
\left[T_{\vec{m}_1},\dots,T_{\vec{m}_{2n}}\right] &= \tfrac{(-2\pi iN)^n}{n!^2}\Big((-1)^n n!^2(2\pi i)^{n}\epsilon^{r_1\dots r_{2n}} \frac{m_{1,r_1}\dots m_{2n,r_{2n}}}{N^n}\\
&\phantom{{}={}}\hspace{5cm}+\CO\left(\tfrac{1}{N^{n+1}}\right)\Big) T_{\vec{m}_1+\dots+\vec{m}_{2n}} \\
&=\Big((2\pi i)^{2n}\epsilon^{r_1\dots r_{2n}} m_{1,r_1}\dots m_{2n,r_{2n}}+\CO\left(\tfrac{1}{N}\right)\Big) T_{\vec{m}_1+\dots+\vec{m}_{2n}}~,
\end{aligned}
\end{equation}
which takes the correct form in the $(N\longrightarrow\infty)$-limit, as desired. For a more detailed proof, see appendix~\ref{app:gen_case_proof}.

\subsubsection{Ordinary matrix representation}
As discussed before, the above generalized algebra in terms of $2n$-tic matrices can equivalently be expressed in terms of ordinary matrices of appropriately larger size. Here, we list the explicit realizations for the general case $p=2n$ making use of the semigroup isomorphism given in~\eqref{eq:gen_semigroup_iso}. The algebra defined by the relations
\begin{equation}
\begin{gathered}
U_k^N=V_k^N= 1~,~~~V_k U_k = \omega U_k V_k~,\\
\left[U_k,U_l\right]=\left[U_k,V_l\right]=\left[V_k,V_l\right]=0~,
\end{gathered}
\end{equation}
with $k,l \in \{1,\dots,n\}~$ can then be realized by the $N^{n}\times N^n$-matrices 
\begin{equation}\label{eq:gen_matrices}
\begin{aligned}
(U_k)_{ij} &= \delta_{\phi_0(i)\phi_0(j)}\dots\delta_{\phi_{n-1}(i)\phi_{n-1}(j)} \omega^{\phi_{k-1}(i)-1}= \delta_{ij} \omega^{\phi_{k-1}(i)-1}~,\\
(V_k)_{ij} &= \delta_{\phi_0(i)\phi_0(j)}\dots\delta^N_{\phi_{k-1}(i)+1,\phi_{k-1}(j)}\dots\delta_{\phi_{n-1}(i)\phi_{n-1}(j)}~,
\end{aligned}
\end{equation}
where $i,j\in \{0\dots,N^n-1\}$, we use the shorthand $\delta^N_{ij} \coloneqq \delta_{i \Mod{N}\,,\,j\Mod{N}}$ and where, again, $\phi_k(x)$ is given by
\begin{equation}
\phi_k(x) = \lfloor\tfrac{x}{N^k}\rfloor\!\! \Mod{N}~.
\end{equation}

These are the images of~\eqref{eq:gen_2n_matrices} under~\eqref{eq:gen_semigroup_iso} and, as $\Phi$ is a semigroup isomorphism, it follows immediately that these matrices satisfy the generalized relations above and, when excluding the identity, now span $\asl(N^n,\FC)$. The assignment of matrices to basis functions and subsequent regularization of the Nambu bracket then follows in the same way as before.

\subsection{Matrix quantization of Nambu brackets for \texorpdfstring{$\mathbf{p\in 2\NN^+-1}$}{p in 2N-1}}
The above Nambu bracket quantization is not suitable for brackets of odd degree $p=2n-1$ further evidencing the even-odd dichotomy discussed in section~\ref{sec:nambu_brackets}. However, we can use the even degree case to define new odd degree brackets by letting one of the entries be constant in a procedure reminiscent of dimensional reduction and familiar from the theory of $n$-Lie algebras, where fixing one of the entries always defines an $(n-1)$-Lie algebra. 

\subsubsection{The cubic case}
We start with the case $n=2~(p=3)$. That is, we have the bracket
\begin{equation}
\left\{\varphi_1,\varphi_2,\varphi_3\right\} = \frac{\epsilon^{rst}}{\sqrt{\omega(\sigma)}}\partial_r\varphi_1\partial_s\varphi_2\partial_t\varphi_3
\end{equation}
and basis functions
\begin{equation}
Y_{\vec{m}} = Y_{(m_1,m_2,m_3)} \coloneqq e^{2\pi i(m_1\sigma_1 + m_2\sigma_2 + m_3\sigma_3)}~,
\end{equation}
resulting in
\begin{equation}\label{eq:cubic_nambu_bracket}
\left\{Y_{\vec{m}},Y_{\vec{n}},Y_{\vec{k}}\right\} =\left(2\pi i\right)^3 \epsilon^{rst} m_r n_s k_t Y_{\vec{m}+\vec{n}+\vec{k}}~,
\end{equation}
which is what we want to approximate.

To motivate using the quartic case to define the finite three-bracket, we first recall that in the infinite-dimensional case we have 
\begin{equation}\label{eq:quartic_nambu_bracket2}
\left\{Y_{\vec{m}'},Y_{\vec{n}'},Y_{\vec{k}'},Y_{\vec{l}'}\right\} =\left(2\pi i\right)^4 \epsilon^{rstu} m'_{r} n'_{s} k'_{t} l'_{u} Y_{\vec{m}'+\vec{n}'+\vec{k}'+\vec{l}'}~,
\end{equation}
where $r,s,t,u=1,2,3,4$. Then, if we let $\vec{m}'=(\vec{m},0)\,,\,\vec{n}'=(\vec{n},0)\,,\,\vec{k}'=(\vec{k},0)$ and lastly $\vec{l}' = (0,0,0,1)$, this leads to
\begin{equation}
\left\{Y_{\vec{m}},Y_{\vec{n}},Y_{\vec{k}},e^{2\pi i \sigma_4}\right\} =\left(2\pi i\right)^4 \epsilon^{\tilde{r}\tilde{s}\tilde{t}} m_{\tilde{r}} n_{\tilde{s}} k_{\tilde{t}} Y_{\vec{m}+\vec{n}+\vec{k}} ~e^{2\pi i \sigma_4}=2\pi i \left\{Y_{\vec{m}},Y_{\vec{n}},Y_{\vec{k}}\right\}e^{2\pi i \sigma_4}~,
\end{equation}
where now $\tilde{r},\tilde{s},\tilde{t}=1,2,3$. This leads us to consider the three-bracket 
\begin{equation}\label{lennarteqn    }
\begin{aligned}
\left[A,B,C\right]&\coloneqq \left[A,B,C,V_2\right] V_2^{-1}\\
&\phantom{:}=\Big(ABCV_2 - ABV_2C - ACBV_2 + ACV_2B + AV_2BC - AV_2CB\\
&\phantom{{}={}} - BACV_2 + BAV_2C + BCAV_2 - BCV_2A - BV_2AC + BV_2CA\\
&\phantom{{}={}} + CABV_2 - CAV_2B - CBAV_2 + CBV_2A + CV_2AB - CV_2BA \\
&\phantom{{}={}}- V_2ABC + V_2ACB + V_2BAC - V_2BCA - V_2CAB + V_2CBA\Big)V_2^{-1}~,
\end{aligned}
\end{equation}
where we use the four-bracket~\eqref{eq:finite_four_bracket} and the (quartic) matrix $V_2$ as defined in~\eqref{eq:quartic_matrix_rep} or~\eqref{eq:gen_matrices}, respectively. Just like the cubic Nambu--Heisenberg commutator, this bracket forms an $L_\infty$-algebra as in~\eqref{eq:odd_linf} when supplemented with the 5-bracket
\begin{equation}
\left[A,B,C,D,E\right] \coloneqq \sum\limits_{i=5}^8\sum\limits_{j=i-4}^{i-1}(\sum\limits_{k=1}^{i-5}+\sum\limits_{k=i+1}^8)\,\, \varepsilon(i,j,k) \left[A,B,C,D,E\right]_{(jk)(i9)}~,
\end{equation}
where $\left[A,B,C,D,E\right]_{(ij)(kl)}$ is the ordinary anti-symmetrized product with $V_2$ and $V_2^{-1}$ inserted in positions $i,j$ and $k,l$, respectively. Here, $\varepsilon(i,j,k)$ is the appropriate minus sign that results from swapping in the $V_2$ and $V_2^{-1}$ into position yielding a completely anti-symmetric 5-bracket as required.

With this in mind we can associate (quartic) matrices to basis functions according to
\begin{equation}
Y_{(m_1,m_2,m_3)} \overset{N\to\infty}{\longleftarrow} \tilde{T}_{(m_1,m_2,m_3)} = \left(\tfrac{i\pi N^2}{2}\right)^{\tfrac{1}{2}} \omega^{\tfrac{m_1m_2}{2}} U_1^{m_1}V_1^{m_2}U_2^{m_3}\sim T_{(m_1,m_2,m_3,0)}~,
\end{equation}
where $U_1,~U_2,~V_1$ are the (quartic) matrices as defined in~\eqref{eq:quartic_matrix_rep} or~\eqref{eq:gen_matrices}, respectively. These (quartic) matrices now satisfy $\tilde{T}^\dagger_{\vec{m}} = i\tilde{T}_{-\vec{m}}$ and span subspaces of $\asl(N^2,\FC)$ or $\quasl(N,\FC)$, respectively. More specifically, the quartic matrices span the subspace 
\begin{equation}
\tensor[_4]{\widetilde{\asl}}{}(N,\FC) \coloneqq \left\{\,A\in\quasl(N,\FC)\mid A_{a_1a_2a_3a_4} = 0 \text{ for } a_2 \neq a_3\,\right\}~,
\end{equation}
while the ordinary matrices span 
\begin{equation}
\widetilde{\asl}(N^2,\FC) \coloneqq \asl(N^2,\FC) \cap \mathsf{BlockDiag}(N,N^2)~,
\end{equation}
where $\mathsf{BlockDiag}(N,N^2)$ are the $N^2\times N^2$-matrices with entries only in the diagonal blocks of size $N\times N$. 

It follows immediately from~\eqref{eq:quartic_basis_product} and~\eqref{eq:quartic_bracket_approx} that
\begin{equation}
\begin{aligned}
\left[\tilde{T}_{\vec{m}},\tilde{T}_{\vec{n}},\tilde{T}_{\vec{k}}\right]&= \tfrac{i\pi N^2}{2}\left(4(2\pi i)^2\epsilon^{rst} \frac{m_r n_s k_t }{N^2} + \CO\left(\tfrac{1}{N^3}\right)\right)\tilde{T}_{\vec{m}+\vec{n}+\vec{k}}\\
&= \Big((2\pi i)^3\epsilon^{rst} m_r n_s k_t  + \CO\left(\tfrac{1}{N}\right)\Big)\tilde{T}_{\vec{m}+\vec{n}+\vec{k}}~,
\end{aligned}
\end{equation}
which in the $N\longrightarrow \infty$ limit recovers~\eqref{eq:cubic_nambu_bracket} as desired.

In order for this definition of a regularized three-bracket to be consistent it needs to satisfy an important consistency check: Repeating the reduction should reproduce the ordinary case of $p=2$, that is the membrane case involving the ordinary commutator and Lie algebra $\asl(N,\FC)$. That is, we define a two-bracket on $\asl(N^2,\FC)\,\,\left(\quasl(N,\FC)\right)$  via
\begin{equation}
\begin{aligned}
&\phantom{\coloneqq}\left[A,B\right]_{\text{red.}}\coloneqq \left[A,B,U_2\right]U_2^{-1}= \left[A,B,U_2,V_2\right]V_2^{-1}U_2^{-1}\\
&=\Big(ABU_2V_2 - ABV_2U_2 - AU_2BV_2 + AU_2V_2B + AV_2BU_2 - AV_2U_2B\\
&\phantom{{}={}} - BAU_2V_2 + BAV_2U_2 + BU_2AV_2 - BU_2V_2A - BV_2AU_2 + BV_2U_2A\\
&\phantom{{}={}} + U_2ABV_2 - U_2AV_2B - U_2BAV_2 + U_2BV_2A + U_2V_2AB - U_2V_2BA \\
&\phantom{{}={}}- V_2ABU_2 + V_2AU_2B + V_2BAU_2 - V_2BU_2A - V_2U_2AB + V_2U_2BA\Big)V_2^{-1}U_2^{-1}~,
\end{aligned}
\end{equation}
where, again, $U_2$ and $V_2$ are the (quartic) matrices defined in~\eqref{eq:quartic_matrix_rep} or~\eqref{eq:gen_matrices}, respectively.

Note that, whenever $A$ and $B$ are linear combinations of basis (quartic) matrices of the form $T_{(m_1,m_2,0,0)}$ only, we can commute $U_2$ and $V_2$ to the right and the bracket simplifies according to
\begin{equation}\label{eq:simplified_reduced_commutator}
\left[A,B\right]_{\text{red}} = 6(1-\omega) \Big(AB- BA\Big)~.
\end{equation}
Then, consider the mapping
\begin{equation}\label{eq:psi_morphism_quartic}
\begin{aligned}
\Psi:&~ \asl(N^2,\FC)\,\,\left(\quasl(N,\FC)\right) \to \asl(N,\FC)~,~~~\\ &~T_{(m_1,m_2,m_3,m_4)} \mapsto 
\begin{cases}
6(1-\omega)(8\pi i N)^{-\tfrac13} T_{(m_1,m_2)}~,&\text{for } (m_3,m_4) = (0,0)~,\\
0~,&\text{for } (m_3,m_4)\in\RZ^2\setminus\left\{(0,0)\right\}~,
\end{cases}
\end{aligned}
\end{equation}
where the basis matrices $T_{(m_1,m_2)}$ on the right correspond to the basis of $\asl(N,\FC)$ generated by the ordinary shift and clock matrices,~cf.~\eqref{eq:membrane_matrix_replacement}. Explicitly, this mapping can be realized as a partial trace 
\begin{equation}\label{eq:explicit_psi1}
\left\{A_{a_1a_2a_3a_4}\right\}_{a_i=0}^{N-1} \overset{\Psi}{\longmapsto} \frac{(i\pi N)^{\tfrac{1}{3}}}{3(1-\omega)}\left\{A_{a_1jja_4}\right\}_{a_1,a_4=0}^{N-1}
\end{equation}
for quartic matrices, where we implicitly sum over the repeated index $j$, and as a summation over diagonal blocks
\begin{equation}\label{eq:explicit_psi2}
\left\{A_{a_1a_2}\right\}_{a_1,a_2=0}^{N^2{-}1} \overset{\Psi}{\longmapsto} \frac{(i\pi N)^{\tfrac{1}{3}}}{3(1-\omega)}\left\{\tfrac{1}{N}\sum\limits_{k=0}^{N-1}A_{i+kN,j+kN}\right\}_{i,j=0}^{N-1}
\end{equation}
for ordinary matrices.

Equipped with this we can consider $\Psi(\left[T_{\vec{m}},T_{\vec{n}}\right]_{\text{red}})$ and, as $\Psi$ vanishes on generators that include $U_2$ or $V_2$, it suffices to consider
\begin{equation}
\begin{aligned}
&\phantom{{}={}}\Psi\Big(\left[T_{(m_1,m_2,0,0)},T_{(n_1,n_2,0,0)}\right]_{\text{red}}\Big) \\
&= \Psi\Big(6(1-\omega)(T_{(m_1,m_2,0,0)}T_{(n_1,n_2,0,0)}-T_{(n_1,n_2,0,0)}T_{(m_1,m_2,0,0)}) \Big)\\
&= \Psi\Big(6(1-\omega)(i\pi N)^{\tfrac23}(\omega^{-\frac{m_1n_2-m_2n_1}{2}}-\omega^{\frac{m_1n_2-m_2n_1}{2}})T_{(m_1+n_1,m_2+n_2,0,0)} \Big)\\
&=36 \cdot 16^{-\tfrac13} (1-\omega)^2 (2i \pi N)^{\tfrac13} (\omega^{-\frac{m_1n_2-m_2n_1}{2}}-\omega^{\frac{m_1n_2-m_2n_1}{2}})T_{(m_1+n_1,m_2+n_2)}\\
&=36 (1-\omega)^2 (8i \pi N)^{-\tfrac23}\left[T_{(m_1,m_2)},T_{(n_1,n_2)}\right]\\
&=\left[\Psi\big(T_{(m_1,m_2,0,0)}\big),\Psi\big(T_{(n_1,n_2,0,0)}\big)\right]~,
\end{aligned}
\end{equation}
where we made use of the simplification~\eqref{eq:simplified_reduced_commutator}. That is, $\Psi$ is a Lie algebra homomorphism and it follows from the first isomorphism theorem that
\begin{equation}
\begin{gathered}
\Big(\,\asl(N^2,\FC)/\ker(\Psi)~,~\left[-,-\right]_{\text{red}}\,\big) \cong \Big(\,\asl(N,\FC)~,~\left[-,-\right] \,\Big) \\
\eand \\
\Big(\,\quasl(N,\FC)/\ker(\Psi)~,~\left[-,-\right]_{\text{red}}\,\Big) \cong \Big(\,\asl(N,\FC)~,~\left[-,-\right] \,\Big)~,
\end{gathered}
\end{equation}
respectively. Note that the quotient spaces $\asl(N^2,\FC)/\ker(\Psi)$ and $\quasl(N,\FC)/\ker(\Psi)$ are the subspaces spanned by the generators of the form $T_{(m_1,m_2,0,0)}\sim T_{(m_1,m_2)}$, so that using the proposed reduction twice does indeed yield a structure isomorphic to the ordinary membrane case.

\subsubsection{The general case}
The above procedure is again generalizable to arbitrary odd degrees $p=2n-1$. That is, in order to approximate the odd degree Nambu bracket
\begin{equation}
\left\{ Y_{\vec{m}_1},\dots,Y_{\vec{m}_{2n-1}}\right\} = (2\pi i)^{2n-1} \epsilon^{r_1\dots r_{2n-1}} m_{1,r_1} \dots m_{2n-1,r_{2n-1}} Y_{\vec{m}_1+\dots +\vec{m}_{2n-1}}~,
\end{equation}
with the basis functions given by
\begin{equation}
Y_{\vec{m}_i} = Y_{(\vec{m}_1,\dots,\vec{m}_{2n-1})} \coloneqq  e^{2\pi i \sum\limits_{j=1}^{2n-1} m_{i,j} \sigma_j}~,
\end{equation}
we define the finite bracket
\begin{equation}
\begin{aligned}
\left[A_1,\dots,A_{2n-1}\right] &\coloneqq \left[A_1,\dots A_{2n-1},V_n\right]V^{-1}_n \\
&\phantom{:}= \sum\limits_{\sigma\in S_{2n}} \mathrm{sgn}(\sigma) A_{\sigma(1)}\dots \hat{A}_{\sigma(j)=2n} V_n\dots A_{\sigma(2n)}V_n^{-1}~,
\end{aligned}
\end{equation}
where $V_n$ is now the appropriate ($2n$-tic) matrix as defined in~\eqref{eq:gen_2n_matrices} or~\eqref{eq:gen_matrices}. Again, just like the higher Nambu--Heisenberg commutators of odd degree, these brackets form an $L_\infty$-algebra of the form~\eqref{eq:odd_linf} when supplemented with an appropriate higher bracket. 

The basis functions are assigned to ($2n$-tic) matrices according to
\begin{equation}
\begin{aligned}
Y_{(m_{i,1},\dots,m_{i,2n-1})} \overset{N\to\infty}{\longleftarrow} \tilde{T}_{(m_{i,1},\dots,m_{i,2n-1})} &= \left(\tfrac{(2\pi i)^{n-1}(-N)^n}{n!^2}\right)^{\tfrac{1}{2n-2}} \omega^{\frac{\sum\limits_{j=1}^{n}m_{i,2j-1}m_{i,2j}}{2}} \\
&\qquad\times U_1^{m_{i,1}}V_1^{m_{i,2}}\dots V_{n-1}^{m_{i,2n-2}}U_n^{m_{i,2n-1}}\\
&\sim T_{(m_{i,1},\dots,m_{i,2n-1},0)}~,
\end{aligned}
\end{equation}
which now span the subspaces
\begin{equation}
\tensor[_{2n}]{\widetilde{\asl}}{}(N,\FC)\coloneqq\left\{\,A\in\quasl(N,\FC)\mid A_{a_1\dots a_n a_{n+1}\dots a_{2n}} = 0 \text{ for } a_n \neq a_{n+1} \,\right\}~,
\end{equation}
and
\begin{equation}
\widetilde{\asl}(N^n,\FC)\coloneqq\asl(N^n,\FC) \cap \mathsf{BlockDiag}(N^{n-1},N^n) ~,
\end{equation}
respectively. It then follows immediately from~\eqref{eq:gen_expanded_bracket} that
\begin{equation}\nonumber
\begin{aligned}
\left[\tilde{T}_{\vec{m}_1},\dots,\tilde{T}_{\vec{m}_{2n-1}}\right]= \Big((2\pi i)^{2n-1}\epsilon^{r_1\dots r_{2n-1}} m_{1,r_1}\dots m_{2n,r_{2n-1}}+\CO\left(\tfrac{1}{N}\right)\Big) \tilde{T}_{\vec{m}_1+\dots+\vec{m}_{2n-1}}~,
\end{aligned}
\end{equation}
as desired.

Again, we check that repeating the above reduction is consistent with the general case of even degree brackets. That is, we have a reduced bracket on both $\asl(N^n,\FC)$ and $\tnasl(N,\FC)$ given by
\begin{equation}\label{eq:gen_reduced_comm}
\left[A_1,\dots,A_{2n-2}\right]_{\text{red}} \coloneqq \left[A_1,\dots,A_{2n-2},U_n,V_n\right]V^{-1}_n U^{-1}_n~,
\end{equation}
where $U_n$ and $V_n$ are, as before, given in~\eqref{eq:gen_2n_matrices} and~\eqref{eq:gen_matrices}, respectively. Then, as a generalization of~\eqref{eq:psi_morphism_quartic}, we consider the mapping
\begin{equation}
\begin{aligned}
\Psi:&~ \asl(N^n,\FC)\,\,\left(\tnasl(N,\FC)\right) \to \asl(N^{n-1},\FC)\,\,\left(\tensor[_{2(n-1)}]{\asl}{}(N,\FC)\right)~,~~~\\ &~T_{(m_{i,1},\dots,m_{i,2n})} \mapsto 
\begin{cases}
\Big((2n^2-n)(1-\omega)\Big)^{\tfrac{1}{2n-3}} &\\
\quad\frac{c(n)}{c(n-1)} T_{(m_{i,1},\dots,m_{i,2n-2})}~,&\text{for } (m_{i,2n-1},m_{i,2n}) = (0,0)~,\\[15pt]
0~,&\text{else },\\
\end{cases}
\end{aligned}
\end{equation}
where $c(n) = ((-2\pi i N)^n/n!^2)^{1/(2n-1)}$. This morphism can be explicitly realized analogously to the expressions given in~\eqref{eq:explicit_psi1} and~\eqref{eq:explicit_psi2}. Again, for matrices not including $U_n$ or $V_n$ the bracket~\eqref{eq:gen_reduced_comm} simplifies as we can commute $U_n$ and $V_n$ to the right and it follows that
\begin{equation}\label{eq:commute_with_Q}
\begin{aligned}
&\phantom{{}={}}\Psi\Big(\left[T_{(m_{1,1},\dots,m_{1,2n-2},0,0)},\dots,T_{(m_{2n-2,1},\dots,m_{2n-2,2n-2},0,0)}\right]_{\text{red}}\Big)\\
&= \Psi\Big(n(2n-1)(1-\omega) \left[T_{(m_{1,1},\dots,m_{1,2n-2},0,0)},\dots,T_{(m_{2n-2,1},\dots,m_{2n-2,2n-2},0,0)}\right]\Big)\\
&= \left[\Psi\big(T_{(m_{1,1},\dots,m_{1,2n-2},0,0)}\big),\dots,\Psi\big(T_{(m_{2n-2,1},\dots,m_{2n-2,2n-2},0,0)}\big)\right]~.
\end{aligned}
\end{equation}
Note that the pairs $\big(\,\asl(N^n,\FC)/\ker(\Psi)\,,\,\left[-,-\right]_{\text{red}}\,\big)$ and $\big(\,\asl(N^{n-1},\FC)\,,\,\left[-,-\right]\,\big)$ form $L_\infty$-algebras $L$ and $L'$ of the form~\eqref{eq:even_l_inf_form}, both with $p=2n-2$. Thus, there is an induced mapping $\mathbf{\Psi}:L\to L'$ where $\Psi$ is applied degree-wise in each of the non-trivial spaces of $L$. In this light, the identity~\eqref{eq:commute_with_Q} is the only requirement in order for $\mathbf{\Psi}$ to be an $L_\infty$-algebra homomorphism, see e.g.~\cite{Lada:1994mn,Saemann:2020sg} for the general definition.

Inversely, we can consider the mapping
\begin{equation}
\begin{aligned}
\Psi^{-1}:&~ \asl(N^{n-1},\FC)\,\,\left(\tensor[_{2(n-1)}]{\asl}{}(N,\FC)\right) \to \asl(N^{n},\FC)\,\,\left(\tnasl(N,\FC)\right)~,~~~\\ &~T_{(m_{i,1},\dots,m_{i,2n-2 })} \mapsto 
\Big((2n^2-n)(1-\omega)\Big)^{2n-3}\frac{c(n-1)}{c(n)} T_{(m_{i,1},\dots,m_{i,2n-2},0,0)}~,
\end{aligned}
\end{equation}
which satisfies the analogous identity to~\eqref{eq:commute_with_Q} and, therefore, induces an $L_\infty$-algebra homomorphism $\mathbf{\Psi}^{-1}:L'\to L$ which is the direct inverse to $\mathbf{\Psi}$. 

That is, $\mathbf{\Psi}$ is an $L_\infty$-algebra isomorphism and it follows that once $\ker\Psi$ is discarded, the structures are isomorphic and the above reduction is indeed self-consistent.

\section{Physical application: a quantized super 4-brane action}\label{sec:four_brane_action}
To provide a physical application, we use the above results to quantize the simple case of a super $(d=9, p=4)$-brane. We start by gauging the action before providing a regularized action and discussing its reduced gauge symmetry.

\subsection{Gauging the super 4-brane action}
We start with the action~\eqref{eq:lc_action_pbrane} which for the super $(d=9, p=4)$-brane takes the form
\begin{equation}\label{gf4}
\begin{aligned}
S_{\mathrm{lc},4-\mathrm{brane}}&= \int \dd\!\tau \dd^4\! \sigma\, \sqrt{\omega(\sigma)} \Big(\frac{P_0^+}{2} \del_0 X^a \del_0 X_a - i P_0^+ \bar\theta \Gamma^- \del_0 \theta\\
&\phantom{{}={}} -\frac{1}{48P_0^+} \left(\left\{X^{a_1},X^{a_2},X^{a_3},X^{a_4}\right\}\right)^2 + i \bar\theta \Gamma^- \Gamma_{abc} \left\{X^a,X^b,X^c,\theta\right\} \Big)~.
\end{aligned}
\end{equation}

To gauge this action we consider the residual symmetries that remain after imposing the lightcone gauge. Similarly to the supermembrane case, these consist of the volume-preserving diffeomorphisms on the $4$-brane. That is, we consider diffeomorphisms of the form  $\sigma^r \rightarrow f^r(\sigma)= \sigma^r + v^r(\sigma)$ with $\partial_{r}v^r=0$ and, as before, use Hodge theory to decompose the solutions to the latter equation into harmonic and co-exact contributions. Again, the only relevant components are the co-exact ones that now take the form
\begin{equation}
v^r(\sigma)=\frac{\epsilon^{rstu}}{\sqrt{w(\sigma)}}\del_s \tilde{\Lambda}_{tu} (\sigma).
\end{equation}

To write the corresponding field transformations in terms of the Nambu brackets, we follow~\cite{Shiba:Diss}. That is, we introduce a trilocal function $\Lambda(\sigma,\sigma',\sigma'')$ by requiring $\tilde{\Lambda}_{tu}(\sigma)\equiv\partial'_t\partial''_u\Lambda(\sigma,\sigma',\sigma'')|_{\sigma=\sigma'=\sigma''}$ which leads to
\begin{equation}\label{tx4}
\begin{aligned}
\delta X^a &= \Lambda_{\vec{b}_1\vec{b}_2\vec{b}_3} \left\{Y^{\vec{b}_1},Y^{\vec{b}_2},Y^{\vec{b}_3},X^a\right\}~,\\
\delta \theta &= \Lambda_{\vec{b}_1\vec{b}_2\vec{b}_3} \left\{Y^{\vec{b}_1},Y^{\vec{b}_2},Y^{\vec{b}_3},\theta\right\}~,
\end{aligned}
\end{equation}
where, $Y^{\vec{b}_i}$ are the toric basis functions~\eqref{ymforlater} and $\Lambda_{\vec{b}_1\vec{b}_2\vec{b}_3}$ are the corresponding basis coefficients of $\Lambda(\sigma,\sigma',\sigma'')$. The action~\eqref{gf4} can directly be seen to indeed remain invariant under these transformations owing to the fundamental identity as well as the identity
\begin{equation}
\int \!\dd^4\sigma\, (\{C,D,E,A\}B+A\{C,D,E,B\})=0~,
\end{equation}
which is a consequence of the Leibniz rule (which turns the
integrand into a sole Nambu bracket) together with integration by parts.

In order to promote these symmetries to local, $\tau$-dependent transformations we again introduce a gauge field $A$. More precisely, we consider a $\tau$-dependent trilocal function $A(\tau,\sigma,\sigma',\sigma'')$ with corresponding basis coefficients $A_{\vec{b}_1\vec{b}_2\vec{b}_3}(\tau)$ that transform under a gauge transformation as
\begin{equation}\label{eq:A_trsf}
\begin{aligned}
\delta A_{\vec{b}_1\vec{b}_2\vec{b}_3}= \partial_0 \Lambda_{\vec{b}_1\vec{b}_2\vec{b}_3} + \Lambda_{\vec{c}_1\vec{c}_2\vec{c}_3} \bigg(&\tensor{f}{^{\vec{c}_1\vec{c}_2\vec{c}_3\vec{d}_{\phantom{1}}}_{\vec{b}_1}}A_{\vec{d}\,\,\vec{b}_2\vec{b}_3}+\tensor{f}{^{\vec{c}_1\vec{c}_2\vec{c}_3\vec{d}_{\phantom{1}}}_{\vec{b}_2}}A_{\vec{b}_1\vec{d\,\,}\vec{b}_3}\\
&+\tensor{f}{^{\vec{c}_1\vec{c}_2\vec{c}_3\vec{d}_{\phantom{1}}}_{\vec{b}_3}}A_{\vec{b}_1\vec{b}_2\vec{d}}\bigg)~,
\end{aligned}
\end{equation}
where $\tensor{f}{^{\vec{b}_1\vec{b}_2\vec{b}_3\vec{b}_4}_{\vec{b}_5}}$ denote the structure constants of the Nambu bracket, i.e. $\{Y^{\vec{b}_1},Y^{\vec{b}_2},Y^{\vec{b}_3},Y^{\vec{b}_4}\}= \tensor{f}{^{\vec{b}_1\vec{b}_2\vec{b}_3\vec{b}_4}_{\vec{b}_5}} Y^{\vec{b}_5}
$. The appropriate covariant derivatives are then defined as
\begin{equation}\label{eq:cov_deriv}
\begin{aligned}
\mathrm{D} X^a &= \partial_0 X^a - A_{\vec{b}_1\vec{b}_2\vec{b}_3}(\tau) \left\{Y^{\vec{b}_1},Y^{\vec{b}_2},Y^{\vec{b}_3},X^a\right\}~,\\
\mathrm{D} \theta &= \partial_0 X^a - A_{\vec{b}_1\vec{b}_2\vec{b}_3}(\tau) \left\{Y^{\vec{b}_1},Y^{\vec{b}_2},Y^{\vec{b}_3},\theta\right\}~,
\end{aligned}
\end{equation}
leading to the gauge invariant action
\begin{equation}\label{gin4}
\begin{aligned}
S_{\mathrm{lc},4-\mathrm{brane}}^{\mathrm{g}} &= \int\! \dd\tau \dd^4 \sigma\, \sqrt{\omega(\sigma)} \Big(\frac{P_0^+}{2} \mathrm{D} X^a \mathrm{D} X_a - i P_0^+ \bar\theta \Gamma^- \mathrm{D} \theta\\
&\phantom{{}={}} -\tfrac{1}{48P_0^+} \left(\left\{X^{a_1},X^{a_2},X^{a_3},X^{a_4}\right\}\right)^2 + i \bar\theta \Gamma^- \Gamma_{abc} \left\{X^a,X^b,X^c,\theta\right\} \Big)~,
\end{aligned}
\end{equation}
which is now invariant under the local versions of the transformations~\eqref{tx4}.

\subsection{Quantizing the super 4-brane action}
In order to quantize the gauge-fixed action~\eqref{gf4}, one would like to replace the quartic Nambu brackets with the approximation by quartic Nambu--Heisenberg commutators found in section~\ref{sec:matrix_approximation}. That is we are interested in the action
\begin{equation}\label{eq:reg_4brane_action}
\begin{aligned}
S_{\mathrm{reg.},4-\mathrm{brane}}&= \int \dd\!\tau\,\, \mathrm{Tr} \Big(\frac{P_0^+}{2} \del_0 X^a \del_0 X_a - i P_0^+ \bar\theta \Gamma^- \del_0 \theta\\
&\phantom{{}={}} -\frac{1}{48P_0^+} \left(\left[X^{a_1},X^{a_2},X^{a_3},X^{a_4}\right]\right)^2 + i \bar\theta \Gamma^- \Gamma_{abc} \left[X^a,X^b,X^c,\theta\right] \Big)~,
\end{aligned}
\end{equation}
which in the $(N\longrightarrow\infty)$-limit is thought to approximate the super $4$-brane action.

However, in attempting to regularize the volume preserving diffeomorphisms~\eqref{tx4}, one finds that the naive operation of also replacing Nambu brackets in these transformations by Nambu--Heisenberg commutators leads to inconsistencies due to the failure to satisfy the fundamental identity and the Leibniz rule. That is, the action~\eqref{eq:reg_4brane_action} is not invariant under the transformations
\begin{equation}\label{tx4f}
\begin{aligned}
\delta X &= \Lambda_{\vec{b}_1\vec{b}_2\vec{b}_3} [T^{\vec{b}_1},T^{\vec{b}_2},T^{\vec{b}_3},X ]~,\\
\delta \theta &= \Lambda_{\vec{b}_1\vec{b}_2\vec{b}_3} [T^{\vec{b}_1},T^{\vec{b}_2},T^{\vec{b}_3},\theta]~,
\end{aligned}
\end{equation}
unlike in the case of the supermembrane. Instead, we introduce the transformations
\begin{equation}\label{eq:red_trafo}
\begin{aligned}
\delta X^a &= \Lambda_{\vec{b}_1\vec{b}_2\vec{b}_3} \bigg([T^{\vec{b}_1},T^{\vec{b}_2},T^{\vec{b}_3}]X^a-X^a[T^{\vec{b}_1},T^{\vec{b}_2},T^{\vec{b}_3}]\bigg)~,\\
\delta \theta &= \Lambda_{\vec{b}_1\vec{b}_2\vec{b}_3} \bigg([T^{\vec{b}_1},T^{\vec{b}_2},T^{\vec{b}_3}]\theta-\theta[T^{\vec{b}_1},T^{\vec{b}_2},T^{\vec{b}_3}]\bigg)~,
    \end{aligned}
\end{equation}
which can be thought of as reduced transformations induced by the bracket
\begin{equation}\label{eq:red_bracket}
\left[A,B,C;D\right]_{\mathrm{red}} = \left[A,B,C\right]D - D \left[A,B,C\right]~.
\end{equation}
Note that --- reminiscent of the generalized $3$-algebras in~\cite{Bagger:2008se} --- this bracket is no longer fully anti-symmetric but rather is only anti-symmetrized in the first three entries. Additionally, it also satisfies neither the fundamental identity nor the Leibniz rule, but nonetheless, a direct computation shows that, due to the cyclicity of the trace, the quantized action~\eqref{eq:reg_4brane_action} is indeed invariant under these transformations.

The action can again be gauged by introducing a gauge field $A$ via a trilocal function $A(\tau,\sigma,\sigma',\sigma'')$ with coefficients $A_{\vec{b}_1\vec{b}_2\vec{b}_3}(\tau)$ that now transform as
\begin{equation}\label{eq:red_gx4}
\begin{aligned}
\delta A_{\vec{b}_1\vec{b}_2\vec{b}_3}= \partial_0 \Lambda_{\vec{b}_1\vec{b}_2\vec{b}_3} + \Lambda_{\vec{c}_1\vec{c}_2\vec{c}_3} \bigg(&\tensor{\tilde{f}}{^{\vec{c}_1\vec{c}_2\vec{c}_3\vec{d}_{\phantom{1}}}_{\vec{b}_1}}A_{\vec{d}\,\,\vec{b}_2\vec{b}_3}+\tensor{\tilde{f}}{^{\vec{c}_1\vec{c}_2\vec{c}_3\vec{d}_{\phantom{1}}}_{\vec{b}_2}}A_{\vec{b}_1\vec{d\,\,}\vec{b}_3}\\
&+\tensor{\tilde{f}}{^{\vec{c}_1\vec{c}_2\vec{c}_3\vec{d}_{\phantom{1}}}_{\vec{b}_3}}A_{\vec{b}_1\vec{b}_2\vec{d}}\bigg)~,
\end{aligned}
\end{equation}
where $\tensor{\tilde{f}}{^{\vec{b}_1\vec{b}_2\vec{b}_3\vec{b}_4}_{\vec{b}_5}}$ are the structure constants corresponding to the reduced bracket, i.e. $\left[T^{\vec{b}_1},T^{\vec{b}_2},T^{\vec{b}_3};T^{\vec{b}_4}\right]_{\mathrm{red}} = \tensor{\tilde{f}}{^{\vec{b}_1\vec{b}_2\vec{b}_3\vec{b}_4}_{\vec{b}_5}} T^{\vec{b}_5}
$. Together with the covariant derivatives
\begin{equation}\label{eq:red_cov_deriv}
\begin{aligned}
\mathrm{D} X^a &= \partial_0 X^a - A_{\vec{b}_1\vec{b}_2\vec{b}_3}(\tau) \left[T^{\vec{b}_1},T^{\vec{b}_2},T^{\vec{b}_3};X^a\right]_{\mathrm{red}}~,\\
\mathrm{D} \theta &= \partial_0 X^a - A_{\vec{b}_1\vec{b}_2\vec{b}_3}(\tau) \left[T^{\vec{b}_1},T^{\vec{b}_2},T^{\vec{b}_3};\theta\right]_{\mathrm{red}}~,
\end{aligned}
\end{equation}
we then arrive at the quantized action
\begin{equation}\label{eq:reg_gauged_4brane_action}
\begin{aligned}
S^{\mathrm{g}}_{\mathrm{reg.},4-\mathrm{brane}}&= \int \dd\!\tau\,\, \mathrm{Tr} \Big(\frac{P_0^+}{2} \mathrm{D} X^a \mathrm{D} X_a - i P_0^+ \bar\theta \Gamma^- \mathrm{D} \theta\\
&\phantom{{}={}} -\frac{1}{48P_0^+} \left(\left[X^{a_1},X^{a_2},X^{a_3},X^{a_4}\right]\right)^2 + i \bar\theta \Gamma^- \Gamma_{abc} \left[X^a,X^b,X^c,\theta\right] \Big)~.
\end{aligned}
\end{equation}

While this action does reduce to the infinite super $4$-brane action in the $A=0$ gauge, it is important to note that the reduced transformations \textit{do not} approximate  the classical gauge transformations~\eqref{tx4} in the large matrix limit. It is unclear whether or not there exist gauge transformations that both constitute symmetries of the quantized action and \textit{do} approximate the classical transformations involving the Nambu bracket. A better understanding of this and related questions would be an interesting topic for future work. Here, we add that both the classical transformations~\eqref{tx4} and the reduced transformations~\eqref{eq:red_gx4} can be understood in terms of the underlying gauge structure induced by the higher brackets.

\subsection{The underlying gauge algebra}
To better understand the above reduced gauge transformations, let us discuss the gauge structure, that is, the $2$-term $L_\infty$-algebra that underlies these transformations. The above transformations~\eqref{tx4} are the $(p=4)$-case of the more general case of Nambu brackets that generate volume-preserving diffeomorphisms in $p$ dimensions.

For $p=3$, the analogous transformations form Bagger--Lambert 3-algebras~\cite{Bagger:2006sk} as they also satisfy the fundamental identity. These have been previously~\cite{Palmer:2012ya,deMedeiros:2008zh,Faulkner:1973aa} shown to be equivalent to  $2$-term $L_\infty$-algebras, or Lie $2$-algebras. As this is the relevant construction we will recall it here.

In our case, we consider the cubic Nambu bracket
\begin{equation}
\left\{X_1,X_2,X_3\right\} = \frac{\epsilon^{rst}}{\sqrt{\omega(\sigma)}}\partial_r X_1 \partial_s X_2 \partial_t X_3~,
\end{equation}
of the functions $X_i \in C^\infty(T^3)$ on the $3$-torus. We define the mapping
\begin{equation}
D: \Omega^2(C^\infty(T^3))\to \mathrm{End}(C^\infty(T^3))~,~~~ D(\Lambda_{\vec{b}_1\vec{b}_2})(X) = \Lambda_{\vec{b}_1\vec{b}_2}\left\{Y^{\vec{b}_1},Y^{\vec{b}_2},X\right\}~,
\end{equation}
where, as before, $Y^{\vec{b}_i}$ are the basis functions as given in~\eqref{ymforlater} with $\sigma_4=0$. Then $\frg = \mathrm{im}(D)$ forms a Lie algebra with the Lie bracket given by the ordinary commutator. This follows directly since the fundamental identity ensures that the ordinary commutator indeed closes within $\mathrm{im}(D)$. 

It subsequently follows that we have a $2$-term $L_\infty$-algebra $(C^\infty(T^3) \overset{0}{\longrightarrow} \frg)~,$ where the non-trivial brackets are given by
\begin{equation}\label{eq:linf_brackets}
\begin{aligned}
\mu_2&: \frg \wedge \frg \to \frg~,~~~\mu_2(D(\Lambda_{\vec{b}_1\vec{b}_2}),D(\Lambda_{\vec{c}_1\vec{c}_2}))(X) = \left[D(\Lambda_{\vec{b}_1\vec{b}_2}),D(\Lambda_{\vec{c}_1\vec{c}_2})\right](X)~,\\
\mu_2&: \frg \wedge C^\infty(T^3) \to C^\infty(T^3)~,~~~  \mu_2(D(\Lambda_{\vec{b}_1\vec{b}_2}),X) = D(\Lambda_{\vec{b}_1\vec{b}_2})(X)~.
\end{aligned}
\end{equation}
The relevant higher Jacobi identities for a $2$-term $L_\infty$-algebra are automatically satisfied as the $\mu_2$ are chosen to be the ordinary commutator and the action of $D$, respectively.

Here, we note that this procedure generalizes to an arbitrary order $p$ of the Nambu bracket resulting in $2$-term $L_\infty$-algebras of the form $(C^\infty(T^p) \overset{0}{\longrightarrow} \frg)$ with the brackets given by the analogous generalizations of~\eqref{eq:linf_brackets}. As we have an $L_\infty$-algebra, the framework for a generalized gauge theory provides immediate expressions for the covariant derivative and corresponding gauge transformations, see e.g.~\cite{Fiorenza:2010mh,Saemann:2015fgj}. In the case of $p=4$, these precisely agree with the expressions in~\eqref{tx4},~\eqref{eq:A_trsf} and~\eqref{eq:cov_deriv}. That is, we have
\begin{equation}
\delta X^a = \mu_2(D(\Lambda_{\vec{b}_1\vec{b}_2\vec{b}_3}),X^a)~,~~~\mathrm{D} X^a = \partial_0 X^a - \mu_2(D(A_{\vec{b}_1\vec{b}_2\vec{b}_3}),X^a)~,
\end{equation}
and
\begin{equation}
\delta D(A_{\vec{b}_1\vec{b}_2\vec{b}_3}) = \partial_0 D(\Lambda_{\vec{c}_1\vec{c}_2\vec{c}_3}) + \mu_2(D(\Lambda_{\vec{c}_1\vec{c}_2\vec{c}_3}),D(A_{\vec{b}_1\vec{b}_2\vec{b}_3}))~.
\end{equation}

In the finite case, the above construction fails as the quartic Nambu--Heisenberg commutator does not satisfy the fundamental identity. More precisely, the commutator of the analogous mapping 
\begin{equation}
D(\Lambda_{\vec{b}_1\vec{b}_2\vec{b}_3})(X) = \Lambda_{\vec{b}_1\vec{b}_2\vec{b}_3} \left[ T^{\vec{b}_1},T^{\vec{b}_2},T^{\vec{b}_3},X \right]
\end{equation}
does not close on the image of $D$ and we therefore cannot form an $L_\infty$-algebra in this way. This can be seen as the underlying reason as to why the gauge transformations induced by the quartic Nambu--Heisenberg commutator do not leave the quantized action~\eqref{gin4} invariant.

However, the reduced bracket~\eqref{eq:red_bracket}, while also not satisfying the fundamental identity, \textit{does} lead to a mapping $D$ whose commutator closes in its image. Thus, this bracket can be used to form an $L_\infty$-algebra whose associated gauge structure agrees with~\eqref{eq:red_trafo},~\eqref{eq:red_gx4} and~\eqref{eq:red_cov_deriv} and \textit{does} leave the quantized action~\eqref{gin4} invariant.

\appendices
\subsection{Explicit calculations for generalized shift and clock matrices}\label{app:explicit_calculations}
To show that the $2n$-tic matrices given in~\eqref{eq:gen_2n_matrices} indeed do satisfy the generalized algebra relations in~\eqref{eq:tntic_shift_and_clock_algebra} we explicitly calculate their products. Note that the analogous result for quartic matrices, that is, the quartic matrices given in~\eqref{eq:quartic_matrix_rep} satisfying the relations~\eqref{eq:quartic_shift_and_clock_algebra} immediately follows from the general case.

For notational clarity we suppress summation signs in the following and implicitly sum over repeated indices. We have
\begin{equation}
\begin{aligned}
(U_i^N)_{a_1\dots a_{2n}} &=\delta_{a_1m_n} \dots \delta_{m_{n(N-2)+1}a_{n+1}} \omega^{N(a_i-1)} =  \delta_{a_1a_{2n}}\dots\delta_{a_n a_{n+1}} = (\tntic{\unit})_{a_1\dots a_{2n}}~,\\
(V_i^N)_{a_1\dots a_{2n}} &= \delta_{a_1a_{2n}}\dots \delta^N_{a_i+N,a_{2n-i+1}}\dots\delta_{a_n a_{n+1}} = \delta_{a_1a_{2n}}\dots\delta_{a_n a_{n+1}} = (\tntic{\unit})_{a_1\dots a_{2n}}~,
\end{aligned}
\end{equation}
as well as
\begin{equation}
\begin{aligned}
(U_i U_j)_{a_1\dots a_{2n}} &= \delta_{a_1m_n} \dots \delta_{a_n m_1} \delta_{m_n a_{2n}} \dots \delta_{m_1 a_{n+1}} \omega^{a_i+m_{n-j+1}-2} \\
&=\delta_{a_1m_n} \dots \delta_{a_n m_1} \delta_{m_n a_{2n}} \dots \delta_{m_1 a_{n+1}} \omega^{m_{n-i+1}+a_j-2} \\
&= (U_j U_i)_{a_1\dots a_{2n}}~,\\
(U_i V_j)_{a_1\dots a_{2n}} &= \delta_{a_1m_n}\dots \delta_{a_n m_1} \delta_{m_n a_{2n}}\dots \delta^N_{m_{n-j+1}+1,a_{2n-j+1}}\dots \delta_{m_1 a_{n+1}} \omega^{a_i-1} \\
&= \delta_{a_1m_n}\dots\delta^N_{a_j+1,m_{n-j+1}}\dots \delta_{a_n m_1} \delta_{m_n a_{2n}}\dots \delta_{m_1 a_{n+1}} \omega^{m_{n-i+1}-1} \\
&= (V_j U_i)_{a_1\dots a_{2n}}~, \\
(V_i V_j)_{a_1\dots a_{2n}} &= \delta_{a_1m_n}\dots\delta^N_{a_i+1,m_{n-i+1}}\dots \delta^N_{m_{n-j+1}+1,a_{2n-j+1}}\dots \delta_{m_1 a_{n+1}} \\
&= \delta_{a_1m_n}\dots\delta^N_{a_j+1,m_{n-j+1}}\dots \delta^N_{m_{n-i+1}+1,a_{2n-i+1}}\dots \delta_{m_1 a_{n+1}}\\
&= (V_j V_i)_{a_1\dots a_{2n}}~,
\end{aligned}
\end{equation}
and, finally,
\begin{equation}
\begin{aligned}
(V_i U_i)_{a_1\dots a_{2n}}&= \delta_{a_1m_n}\dots \delta^N_{a_i+1,m_{n-i+1}}\dots \delta_{a_n m_1} \delta_{m_n a_{2n}}\dots \delta_{m_1 a_{n+1}} \omega^{m_{n-i+1}-1}\\
&=\omega\left(\delta_{a_1m_n}\dots \delta_{a_n m_1} \delta_{m_n a_{2n}}\dots \delta^N_{m_{n-i+1}+1,a_{2n-i+1}}\dots \delta_{m_1 a_{n+1}} \omega^{a_i-1}\right) \\
&=\omega (U_i V_i)_{a_1\dots a_{2n}}~.
\end{aligned}
\end{equation}

Lastly, it also follows that the ordinary matrices in~\eqref{eq:gen_matrices} satisfy the generalized relations~\eqref{eq:tntic_shift_and_clock_algebra} as they are the images of the $2n$-tic matrices~\eqref{eq:gen_2n_matrices} under the semigroup isomorphism~\eqref{eq:gen_semigroup_iso}.

\subsection{Proof of finite bracket regularization in the general case}\label{app:gen_case_proof}
In order to prove the general expansion as in~\eqref{eq:gen_expanded_bracket} we need to prove that
\begin{equation}\label{eq:exponent_alt_sum_formula}
\sum\limits_{\sigma\in S_{2n}} \mathrm{sgn}(\sigma) e^k_{2n}(\vec{m}_{\sigma(1)},\dots,\vec{m}_{\sigma(2n)}) = 
\begin{cases}
0~,& k < n~,\\
(-1)^n n!^2\epsilon^{r_1\dots r_{2n}} m_{1,r_1}\dots m_{2n,r_{2n}}~,& k = n~,
\end{cases}
\end{equation}
where $e_{2n}(\vec{m}_{\sigma(1)},\dots,\vec{m}_{\sigma(2n)})$ as in~\eqref{eq:gen_exponent} is given by
\begin{equation}\label{eq:app_e}
e_{2n}(\vec{m}_1,\dots,\vec{m}_{2n}) = \frac12\sum\limits_{j=1}^n \sum\limits_{i=1}^{2n}m_{i,2j-1}\left(\sum\limits_{k=1}^{i-1}m_{k,2j} - \sum\limits_{k=i+1}^{2n}m_{k,2j}\right)~.
\end{equation}

In order to prove this, we will examine the terms appearing on a case-by-case basis. Note that $e^k_{2n}(\vec{m}_1,\dots,\vec{m}_{2n})$ is a homogeneous polynomial of degree $2k$ in the vector entries $m_{i,j}$. Let us denote the terms appearing by
\begin{equation}
t(\vec{i},\vec{j}) = m_{i_1,j_1}\dots m_{i_{2k},j_{2k}}~,
\end{equation}
and let us also denote the cardinality of the set consisting of the entries of the vectors $\vec{i}$ and $\vec{j}$ by $|\{\,\vec{i}\,\}|$ and $|\{\,\vec{j}\,\}|$, respectively. That is, we have $|\{\,\vec{i}\,\}|\leq 2k$ and $|\{\,\vec{j}\,\}|\leq 2k$. The formal vector space spanned by the terms $t(\vec{i},\vec{j})$ forms a representation of the symmetric group $S_{2n}$ where a permutation acts via
\begin{equation}
\sigma\acton t(\vec{i},\vec{j}) = m_{\sigma(i_1),\,j_1}\dots m_{\sigma(i_{2k}),\,j_{2k}}~.
\end{equation}
In~\eqref{eq:exponent_alt_sum_formula}, the alternating sum projects this representation onto the alternating representation of $S_{2n}$ and, thus, we only have to consider the terms $t(\vec{i},\vec{j})$ which can possibly form an alternating representation. The values of $k,\,|\{\,\vec{i}\,\}|$ and $|\{\,\vec{j}\,\}|$ fully enumerate the possibilities, which we will discuss in order.

First, it is immediate that there cannot be any alternating representations if $k<n$. That is, when $k<n$ we necessarily have $|\{\,\vec{i}\,\}|\leq 2k  \leq 2n-2$ for all terms $t(\vec{i},\vec{j})$ and, thus, each $t(\vec{i},\vec{j})$ in $e^k_{2n}(\vec{m}_1,\dots,\vec{m}_{2n})$ is independent of at least two of the vectors $\vec{m}_l$ and thus symmetric under their exchange. Therefore, it follows that the alternating sum in~\eqref{eq:exponent_alt_sum_formula} vanishes for $k<n$.

Second, for $k=n$ there are $3n^2{-}3n-1$ different potential realizations of the alternating representation of which only one is realized in $e^n_{2n}(\vec{m}_{\sigma(1)},\dots,\vec{m}_{\sigma(2n)})$:

Again, for terms $t(\vec{i},\vec{j})$ with $|\{\,\vec{i}\,\}| \leq 2n-2$ the alternating sum necessarily vanishes, so we only need to consider terms with $|\{\,\vec{i}\,\}| = 2n-1$ or $|\{\,\vec{i}\,\}| = 2n$.

Starting with terms $t(\vec{i},\vec{j})$ where $|\{\,\vec{i}\,\}| = 2n-1$, we can discard terms with $|\{\,\vec{j}\,\}| < 2n-2$, as those terms have to include at least one pair of the form $m_{i,j}m_{i',j},~i \neq i'$ with both $i$ and $i'$ only appearing once in $\vec{i}$. Thus, they become symmetric under the exchange $i\leftrightarrow i'$ and, again, vanish in the alternating sum.

Then, as the structure of $e_{2n}(\vec{m}_1,\dots,\vec{m}_{2n})$ requires that $|\{\,\vec{j}\,\}|$ is even, the remaining terms with $|\{\,\vec{i}\,\}|=2n-1$ either have $|\{\,\vec{j}\,\}| = 2n-2$ or $|\{\,\vec{j}\,\}| = 2n$. More explicitly, the only relevant terms are of the form
\begin{equation}
\begin{gathered}
t_1(\vec{i},\vec{j})=m_{i,j}m_{i,j+1}m_{i',j}m_{i'',j+1}\dots~,~j\in 2\RZ+1~,\\
|\{\,\vec{i}\,\}| = 2n-1~~\mathrm{and}~~|\{\,\vec{j}\,\}| = 2n-2~,
\end{gathered}
\end{equation}
of which there are $n(n-1)$ many different realizations, and
\begin{equation}
\begin{gathered}
t_2(\vec{i},\vec{j})=m_{i,j}m_{i,j'}m_{i',j''}m_{i'',j'''}\dots~,\lfloor \tfrac{j-1}{2}\rfloor \neq \lfloor\tfrac{j'-1}{2}\rfloor~,~|j-j''|=|j'-j'''|=1~,\\
|\{\,\vec{i}\,\}| = 2n-1~~\mathrm{and}~~ |\{\,\vec{j}\,\}| = 2n~,
\end{gathered}
\end{equation}
of which there are $2n(n-1)$ many. It is now straightforward to see that in the multinomial expansion of $e^{n}_{2n}(\vec{m}_1,\dots,\vec{m}_{2n})$ each term of the types $t_1(\vec{i},\vec{j})$ and $t_2(\vec{i},\vec{j})$ has a unique way to be made up out of individual factors and, thus, appears with the same multinomial coefficient. Alternating sums of such terms do not vanish and can thus form alternate representations. We will, however, see in the following that such combinations do not appear in $e^n_{2n}(\vec{m}_1,\dots,\vec{m}_{2n})$.

Starting with $t_1(\vec{i},\vec{j})$ let us assume that $i<\min(i',i'')$ and consider the term $t^\ast_1(\vec{i},\vec{j})$ defined to be given by $t_1(\vec{i},\vec{j})$ but with the role of $i$ swapped with $\max(i',i'')$. As $t^\ast_1(\vec{i},\vec{j})$ is of the same form as $t_1(\vec{i},\vec{j})$, it necessarily also appears in the multinomial expansion of $e^{n}_{2n}(\vec{m}_1,\dots,\vec{m}_{2n})$ with the same multinomial coefficient.

Additionally, both $t_1^\ast(\vec{i},\vec{j})$ and $t_1(\vec{i},\vec{j})$ appear with the same relative sign in the expansion of $e^n_{2n}(\vec{m}_1,\dots,\vec{m}_{2n})$ which can be seen as follows: In $e_{2n}(\vec{m}_1,\dots,\vec{m}_{2n})$ a generic term $m_{i_1,j}m_{i_2,j+1}$ appears with a plus sign if $i_1<i_2$ and with a minus sign if $i_1>i_2$. Thus, the difference in sign between $t_1(\vec{i},\vec{j})$ and $t^\ast_1(\vec{i},\vec{j})$ is solely due to the change in signs between $m_{i,j+1}m_{i',j}$ and $m_{i',j+1}m_{i,j}$ as well as $m_{i,j}m_{i'',j+1}$ and $m_{i',j}m_{i'',j+1}$ if $i'>i''$, while if $i''>i'$ it is solely due to the change in signs between $m_{i,j+1}m_{i',j}$ and $m_{i'',j+1}m_{i',j}$ as well as $m_{i,j}m_{i'',j+1}$ and $m_{i'',j}m_{i,j+1}$. In either case, both change sign and, thus, cancel each other out so that $t_1(\vec{i},\vec{j})$ and $t^\ast_1(\vec{i},\vec{j})$ do indeed appear with the same relative sign. 

That is, in the expansion of $e^{n}_{2n}(\vec{m}_1,\dots,\vec{m}_{2n})$ the terms $t_1(\vec{i},\vec{j})$ with $i<\min(i',i'')$ are paired with terms $t_1^\ast(\vec{i},\vec{j})$ which are the terms with $i>\min(i',i'')$. As these have the same prefactor, the sum $t_1(\vec{i},\vec{j})+t_1^\ast(\vec{i},\vec{j})$ is symmetric under the exchange $i\leftrightarrow \max(i',i'')$ and therefore vanish in the alternating sum of~\eqref{eq:exponent_alt_sum_formula}. 

Similarly, for $t_2(\vec{i},\vec{j})$ with $i<\min(i,i'')$ we construct $t^\ast_2(\vec{i},\vec{j})$ by exchanging the role of $i$ with $\max(i',i'')$. As before, $t^\ast_2(\vec{i},\vec{j})$ consists of the terms where $i>\min(i',i'')$ so that every term has a partner. By an analogous argument to the above, both $t_2(\vec{i},\vec{j})$ and $t^\ast_2(\vec{i},\vec{j})$ appear with the same multinomial coefficient and relative sign in the expansion of $e^{n}_{2n}(\vec{m}_1,\dots,\vec{m}_{2n})$ and thus form pairs that are symmetric under the exchange of $i$ with $\max(i',i'')$. Thus, again, they vanish in the alternating sum of~\eqref{eq:exponent_alt_sum_formula}.

Lastly, we have terms with $|\{\,\vec{i}\,\}|=2n$. Similarly to the above, we can now discard terms with $|\{\,\vec{j}\,\}|<2n$ as these necessarily contain at least one pair of the form $m_{i,j}m_{i',j},~i \neq i'$ with both $i$ and $i'$ appearing uniquely in $\vec{i}$ making them symmetric under the exchange $i\leftrightarrow i'$. Therefore, there is only one possible realization of the alternating representation that appears in $e^n_{2n}(\vec{m}_{\sigma(1)},\dots,\vec{m}_{\sigma(2n)})$ given by the terms with $|\{\,\vec{i}\,\}|=|\{\,\vec{j}\,\}|=2n$ and we will see that this realization indeed appears. Note that, again, there is a unique way for these terms to be made up out of individual factors in the multinomial expansion of $e^{n}_{2n}(\vec{m}_1,\dots,\vec{m}_{2n})$ so that they all appear with the same multinomial coefficient $\pm n!/2^n$.

More specifically, let us consider the term $m_{1,1}\dots m_{2n,2n}$ and count how many times it appears in the sum of~\eqref{eq:exponent_alt_sum_formula}. For a given $\sigma\in S_{2n}$ the structure of the exponent~\eqref{eq:app_e} implies that each pair of the form $m_{i,i}m_{i+1,i+1}$, with $i\in 2\RZ-1$, contributes a minus sign when $\sigma^{-1}(i)<\sigma^{-1}(i+1)$ and a plus sign when $\sigma^{-1}(i)>\sigma^{-1}(i+1)$ to the multinomial expansion. It follows that if there exists an $i^\ast\in 2\RZ-1$ such that $\lfloor\tfrac{\sigma(i^\ast)-1}{2}\rfloor\neq\lfloor\tfrac{\sigma(i^\ast+1)-1}{2}\rfloor$, then $m_{1,1}\dots m_{2n,2n}$ must appear with the same sign in both $e^n_{2n}(\vec{m}_{\sigma(1)},\dots,\vec{m}_{\sigma(2n)})$ and $e^n_{2n}(\vec{m}_{\rho(1)},\dots,\vec{m}_{\rho(2n)})$ where $\rho = (i^\ast\,\,i^\ast+1)\circ\sigma$. Indeed, as $\lfloor\tfrac{\sigma(i^\ast)-1}{2}\rfloor\neq\lfloor\tfrac{\sigma(i^\ast+1)-1}{2}\rfloor$ we have that transposing the position of $\vec{m}_{i^\ast}$ and $\vec{m}_{i^\ast+1}$ does not change any of the relevant order. However, as $\sigma$ and $\rho$ have opposite degrees it follows that these two contributions to $m_{1,1}\dots m_{2n,2n}$ must cancel in the alternating sum~\eqref{eq:exponent_alt_sum_formula}.

That is, we are left with permutations $\sigma\in S_{2n}$ where no such $i^\ast$ exist. Here the opposite is true, as any transposition $(i \,\, i+1)$, where $i\in 2\RZ-1$, must induce a minus sign and thus this contribution does not vanish. There are $n!$ many ways to order the $n$ pairs $(i,i+1)$ and there are two ways of ordering each of those pairs, so that we have $n!2^n$ many copies of $\pm n!/2^n m_{1,1}\dots m_{2n,2n}$ in the alternating sum. The same arguments hold for terms of the form $m_{\sigma(1),1}\dots m_{\sigma(2n),2n}$, when we account for the fact that $\sigma$ just defines a new relative order.

Lastly, note that it is immediate from~\eqref{eq:app_e} that $m_{1,1}\dots m_{2n,2n}$ appears with a factor of $(-1)^n n!/2^n$ in $e^n_{2n}(\vec{m_1},\dots,\vec{m_{2n}})$ and we altogether arrive at
\begin{equation}
\begin{aligned}
\sum\limits_{\sigma\in S_{2n}} \mathrm{sgn}(\sigma) e^n_{2n}(\vec{m}_{\sigma(1)},\dots,\vec{m}_{\sigma(2n)}) &= 2^n n!\left((-1)^n \tfrac{n!}{2^n} \epsilon^{r_1\dots r_{2n}} m_{1,r_1}\dots m_{2n,r_2n}\right)\\
&= (-1)^n n!^2\epsilon^{r_1\dots r_{2n}} m_{1,r_1}\dots m_{2n,r_2n}~.
\end{aligned}
\end{equation}

\end{document}